\documentclass[12pt,letterpaper,english,aps,superscriptaddress,groupedaddress,showpacs,showkeys]{revtex4}
\usepackage[T1]{fontenc}
\usepackage[latin9]{inputenc}
\usepackage{babel}
\usepackage{amsmath}
\usepackage{amssymb}
\usepackage{esint}
\usepackage[unicode=true]
 {hyperref}
\usepackage{breakurl}

\makeatletter

\@ifundefined{textcolor}{}
{%
 \definecolor{BLACK}{gray}{0}
 \definecolor{WHITE}{gray}{1}
 \definecolor{RED}{rgb}{1,0,0}
 \definecolor{GREEN}{rgb}{0,1,0}
 \definecolor{BLUE}{rgb}{0,0,1}
 \definecolor{CYAN}{cmyk}{1,0,0,0}
 \definecolor{MAGENTA}{cmyk}{0,1,0,0}
 \definecolor{YELLOW}{cmyk}{0,0,1,0}
}

\usepackage{dcolumn}
\usepackage{bm}
\usepackage{amsfonts}

\makeatother

\begin{document}

\title{On 3D and 1D Weyl particles in a 1D box }

\author{Salvatore De Vincenzo}

\homepage{https://orcid.org/0000-0002-5009-053X}

\email{[salvatore.devincenzo@ucv.ve]}

\selectlanguage{english}%

\affiliation{Escuela de F\'{\i}sica, Facultad de Ciencias, Universidad Central de Venezuela,
A.P. 47145, Caracas 1041-A, Venezuela.}

\thanks{I would like to dedicate this paper to the memory of my beloved father
Carmine De Vincenzo Di Fresca, who passed away unexpectedly on March
16, 2018. That day something inside of me also died.}

\date{August 29, 2020}
\begin{abstract}
\noindent \textbf{Abstract} We construct the most general families
of self-adjoint boundary conditions for three (equivalent) Weyl Hamiltonian
operators, each describing a three-dimensional Weyl particle in a
one-dimensional box situated along a Cartesian axis. These results
are essentially obtained by using the most general family of self-adjoint
boundary conditions for a Dirac Hamiltonian operator that describes
a one-dimensional Dirac particle in a box, in the Weyl representation,
and by applying simple changes of representation to this operator.
Likewise, we present the most general family of self-adjoint boundary
conditions for a Weyl Hamiltonian operator that describes a one-dimensional
Weyl particle in a one-dimensional box. We also obtain and discuss
throughout the article distinct results related to the Weyl equations
in (3+1) and (1+1) dimensions, in addition to their respective wave
functions, and present certain key results related to representations
for the Dirac equation in (1+1) dimensions.
\end{abstract}



\maketitle

\section{Introduction}

\noindent In 1928, in the first edition of his book in German \cite{RefA},
and in 1929, in a couple of articles \cite{RefB,RefC}, Weyl proposed
-- among other important things -- two two-component wave equations
for the description of free massless fermions in (3+1) dimensions
\cite{RefD,RefE} (an English translation of Ref. \cite{RefC} can
be seen in Ref. \cite{RefF}). In 1957, Lee and Yang chose to assign
one of these two equations specifically to the neutrino \cite{RefG},
but in 1958, Feynman and Gell-Mann showed that it was actually correct
to assign the other equation to this particle \cite{RefH}. Because
there is now experimental evidence that a neutrino has a very small
rest mass, the Weyl equations only approximately describe the behavior
of this particle and its antiparticle. In passing, the (free) Weyl
equations admit the standard minimal substitution and therefore admit
an external electromagnetic four-potential; thus, these equations
could also approximately describe the behavior of charged light fermions.
In general, it can be said that the Weyl equations in (3+1) dimensions
describe three-dimensional Weyl particles (i.e., 3D Weyl particles),
and in (1+1) dimensions, they describe one-dimensional Weyl particles
(i.e., 1D Weyl particles).

The Weyl equations in (3+1) and (1+1) dimensions are more easily constructed
from the respective Dirac equation (in its respective Weyl representation).
A particularly nice derivation of these equations in (3+1) dimensions
can be seen in Ref. \cite{RefI}, p. 79. In this reference, the Weyl
equations were obtained by linearizing the van der Waerden second-order
equation and then making the rest mass of the particle zero. Other
derivations can be seen, for example, in Refs. \cite{RefJ,RefK,RefL,RefM}. 

Among other things, we want to explicitly obtain the most general
family of self-adjoint boundary conditions for each of the three (equivalent)
one-dimensional Cartesian reductions of the (three-dimensional) Weyl
Hamiltonian operator (i.e., the usually named Dirac-Weyl Hamiltonian
operator) for a 3D Weyl particle inside a three-dimensional square
box. Each of these three operators describes a 3D Weyl particle that
is ultimately restricted to a one-dimensional box of width $\ell$
situated along a Cartesian axis. Essentially, we obtain these three
families of boundary conditions from the most general family of self-adjoint
boundary conditions for a Dirac Hamiltonian operator that describes
a 1D Dirac particle in a box of width $\ell$, in the Weyl representation.
That is, we can obtain the families of boundary conditions for each
of the three Weyl Hamiltonian operators by using only the aforementioned
(Dirac) general family of boundary conditions and making some simple
changes of representation. All of these results are presented and
developed in section III. Key results pertaining to representations
for the Dirac equation in (1+1) dimensions are also presented in this
section. Before this, in section II, we present distinct results related
to the Weyl equations in (3+1) and (1+1) dimensions and their respective
wave functions. In section IV, we present the most general family
of self-adjoint boundary conditions for a Weyl Hamiltonian operator
that describes a 1D Weyl particle, also in a one-dimensional box of
width $\ell$. Finally, our conclusions are presented in section V,
and some results that complement what has been stated throughout the
article are exhibited in the appendix.

\section{Results pertaining to the 3D and 1D Weyl equations}

\subsection{In (3+1) dimensions}

\noindent The equation for a single free massless Dirac particle in
(3+1) dimensions has the form
\begin{equation}
\mathrm{i}\hat{\gamma}^{\mu}\partial_{\mu}\Psi=0,
\end{equation}
where $\Psi$ is the four-component Dirac wave function (or Dirac
spinor), $\partial_{\mu}=(c^{-1}\partial_{t},\nabla)$, and the Dirac
gamma matrices $\hat{\gamma}^{\mu}$, with $\mu=0,j$, and $j=1,2,3$,
satisfy the Clifford relation $\hat{\gamma}^{\mu}\hat{\gamma}^{\nu}+\hat{\gamma}^{\nu}\hat{\gamma}^{\mu}=2g^{\mu\nu}\hat{1}_{4}$,
where $g^{\mu\nu}=\mathrm{diag}(1,-1,-1,-1)$ is the metric tensor
($\hat{1}_{4}$ is the $4\times4$ identity matrix). Additionally,
we have the relation $(\hat{\gamma}^{\mu})^{\dagger}=\hat{\gamma}^{0}\hat{\gamma}^{\mu}\hat{\gamma}^{0}$
($\dagger$ denotes the Hermitian conjugate, or the adjoint, of a
matrix and an operator, as usual). In the Weyl representation (or
the chiral or spinor representation), the four-component wave function
and the Dirac gamma matrices can be written as follows \cite{RefN}:
\begin{equation}
\Psi\equiv\left[\begin{array}{c}
\varphi_{1}\\
\varphi_{2}
\end{array}\right]\,,\quad\hat{\gamma}^{\mu}=\left[\begin{array}{cc}
\hat{0}_{2} & -\hat{\bar{\sigma}}^{\mu}\\
-\hat{\sigma}^{\mu} & \hat{0}_{2}
\end{array}\right].
\end{equation}
Likewise, the top two-component wave function can be written as $\varphi_{1}\equiv\left[\,\varphi_{1}^{\mathrm{t}}\;\varphi_{1}^{\mathrm{b}}\,\right]^{\mathrm{T}}$
and the bottom one as $\varphi_{2}\equiv\left[\,\varphi_{2}^{\mathrm{t}}\;\varphi_{2}^{\mathrm{b}}\,\right]^{\mathrm{T}}$,
where $\varphi_{1}^{\mathrm{t}}$ and $\varphi_{2}^{\mathrm{t}}$
are the top components and $\varphi_{1}^{\mathrm{b}}$ and $\varphi_{2}^{\mathrm{b}}$
are the bottom components of the respective two-component wave function
($\mathrm{^{T}}$ represents the transpose of a matrix). In (3+1)
dimensions, $\Psi$ is usually called the bispinor. Additionally,
we have $\hat{\sigma}^{\mu}\equiv(\hat{1}_{2},\hat{\sigma}_{x},\hat{\sigma}_{y},\hat{\sigma}_{z})\equiv(\hat{1}_{2},\hat{\boldsymbol{\sigma}})$
and $\hat{\bar{\sigma}}^{\mu}\equiv(\hat{1}_{2},-\hat{\sigma}_{x},-\hat{\sigma}_{y},-\hat{\sigma}_{z})\equiv(\hat{1}_{2},-\hat{\boldsymbol{\sigma}})$
($\hat{1}_{2}$ is the $2\times2$ identity matrix and $\hat{\sigma}_{j}$'s
are the usual Pauli matrices). Additionally, in Eq. (2), $\hat{0}_{2}$
is the $2$-dimensional zero matrix. 

By substituting $\Psi$ and $\hat{\gamma}^{\mu}$ from Eq. (2) into
Eq. (1), we derive the well-known (explicitly covariant) two-component
(free) Weyl equations, namely, 
\begin{equation}
\mathrm{i}\hat{\sigma}^{\mu}\partial_{\mu}\varphi_{1}=0
\end{equation}
and 
\begin{equation}
\mathrm{i}\hat{\bar{\sigma}}^{\mu}\partial_{\mu}\varphi_{2}=0,
\end{equation}
where $\varphi_{1}$ and $\varphi_{2}$ are called Weyl spinors. Let
us now forget how we obtained these two equations. As is known, Eq.
(3) is usually assigned to the massless antineutrino and Eq. (4) to
the massless neutrino \cite{RefJ}. However, it is possible that Eq.
(4) only (or Eq. (3) only) is sufficient for the description of a
massless fermion, a case called Weyl's two-component theory \cite{RefI}.
Moreover, Eqs. (3) and (4) are non-equivalent two-component equations
in the sense that $\varphi_{1}$ and $\varphi_{2}$ transform in two
different ways under a Lorentz boost, i.e., they transform according
to two inequivalent representations of the Lorentz group \cite{RefM}
(although these two two-component wave functions transform in the
same way under rotations). In fact, let us write the typical Lorentz
boost with speed $v$ along the $x^{j}$-axis in the following way:
$\left[\, ct'\; x'\; y'\; z'\,\right]^{\mathrm{T}}=\hat{\Lambda}_{j}\left[\, ct\; x\; y\; z\,\right]^{\mathrm{T}}$
(i.e., $x^{\mu}\,'=(\Lambda_{j})_{\;\,\nu}^{\mu}\, x^{\nu}$$\,\Rightarrow x^{\mu}=(\Lambda_{j}^{-1})_{\;\,\nu}^{\mu}\, x^{\nu}\,'$),
where
\begin{equation}
\hat{\Lambda}_{j}=\hat{\Lambda}_{j}(\omega)=\exp\left(\mathrm{i}\omega\hat{K}_{j}\right),
\end{equation}
with 
\begin{equation}
\hat{K}_{1}=\left[\begin{array}{cc}
\mathrm{i}\hat{\sigma}_{x} & \hat{0}_{2}\\
\hat{0}_{2} & \hat{0}_{2}
\end{array}\right]\,,\quad\hat{K}_{2}=\left[\begin{array}{cc}
\hat{0}_{2} & \frac{\mathrm{i}}{2}(\hat{1}_{2}+\hat{\sigma}_{z})\\
\frac{\mathrm{i}}{2}(\hat{1}_{2}+\hat{\sigma}_{z}) & \hat{0}_{2}
\end{array}\right]\,,\quad\hat{K}_{3}=\left[\begin{array}{cc}
\hat{0}_{2} & \frac{\mathrm{i}}{2}(\hat{\sigma}_{x}+\mathrm{i}\hat{\sigma}_{y})\\
\frac{\mathrm{i}}{2}(\hat{\sigma}_{x}-\mathrm{i}\hat{\sigma}_{y}) & \hat{0}_{2}
\end{array}\right],
\end{equation}
and, as usual, $\tanh(\omega)=v/c\equiv\beta$ and $\cosh(\omega)=(1-\beta^{2})^{-1/2}\equiv\gamma$
(the speed of the primed reference frame in the direction of the $x^{j}$-axis
with respect to the unprimed reference frame is precisely $v$). Then,
under this Lorentz boost, the Dirac wave function in (3+1) dimensions
transforms as $\Psi'(x^{j}\,',t')=\hat{S}(\Lambda_{j})\Psi(x^{j},t)$,
where $\hat{S}(\Lambda_{j})=\exp(-\omega\hat{\gamma}^{0}\hat{\gamma}^{j}/2)$
and the latter matrix obeys the relation $(\Lambda_{j})_{\;\,\nu}^{\mu}\hat{\gamma}^{\nu}=\hat{S}^{-1}(\Lambda_{j})\,\hat{\gamma}^{\mu}\,\hat{S}(\Lambda_{j})$.
Then, the $4\times4$ matrix $\hat{S}(\Lambda_{j})$ is a block-diagonal
matrix (as expected in the Weyl representation), and we obtain the
following results \cite{RefO}:
\begin{equation}
\left[\begin{array}{c}
\varphi_{1}^{\mathrm{t}}\,'(x^{j}\,',t')\\
\varphi_{1}^{\mathrm{b}}\,'(x^{j}\,',t')
\end{array}\right]=\left[\,\cosh\left(\frac{\omega}{2}\right)\hat{1}_{2}-\sinh\left(\frac{\omega}{2}\right)\hat{\sigma}_{j}\,\right]\left[\begin{array}{c}
\varphi_{1}^{\mathrm{t}}(x^{j},t)\\
\varphi_{1}^{\mathrm{b}}(x^{j},t)
\end{array}\right]
\end{equation}
and 
\begin{equation}
\left[\begin{array}{c}
\varphi_{2}^{\mathrm{t}}\,'(x^{j}\,',t')\\
\varphi_{2}^{\mathrm{b}}\,'(x^{j}\,',t')
\end{array}\right]=\left[\,\cosh\left(\frac{\omega}{2}\right)\hat{1}_{2}+\sinh\left(\frac{\omega}{2}\right)\hat{\sigma}_{j}\,\right]\left[\begin{array}{c}
\varphi_{2}^{\mathrm{t}}(x^{j},t)\\
\varphi_{2}^{\mathrm{b}}(x^{j},t)
\end{array}\right].
\end{equation}
Thus, we have two different kinds of two-component Weyl wave functions
(or Weyl spinors) in (3+1) dimensions. That is, the result in Eq.
(7) is for a type of 3D Weyl particle, for example, for a massless
antineutrino, and that in Eq. (8) is for the other type of 3D Weyl
particle, for example, for a massless neutrino. 

Usually, Eq. (3) is called the right-chiral Weyl equation and Eq.
(4) is called the left-chiral Weyl equation because the four-component
Dirac wave functions $\Psi_{+}=\left[\,\varphi_{1}\;0\,\right]^{\mathrm{T}}=\frac{1}{2}(\hat{1}_{4}+\hat{\gamma}^{5})\Psi$
and $\Psi_{-}=\left[\,0\;\varphi_{2}\,\right]^{\mathrm{T}}=\frac{1}{2}(\hat{1}_{4}-\hat{\gamma}^{5})\Psi$
are eigenstates of the so-called chirality matrix $\hat{\gamma}^{5}\equiv\mathrm{i}\hat{\gamma}^{0}\hat{\gamma}^{1}\hat{\gamma}^{2}\hat{\gamma}^{3}=\mathrm{diag}(\hat{1}_{2},-\hat{1}_{2})$,
namely, $\hat{\gamma}^{5}\Psi_{\pm}=(\pm1)\Psi_{\pm}$ ($\Psi_{+}$
is the right-chiral eigenstate and $\Psi_{-}$ is the left-chiral
eigenstate). The chirality matrix is Hermitian and satisfies the relations
$(\hat{\gamma}^{5})^{2}=\hat{1}_{4}$, and $\hat{\gamma}^{5}\hat{\gamma}^{\mu}+\hat{\gamma}^{\mu}\hat{\gamma}^{5}=\mathrm{diag}(\hat{0}_{2},\hat{0}_{2})$
\cite{RefO,RefP}. Incidentally, because the matrices $\hat{S}(\Lambda_{j})$
-- boosts and rotations -- commute with $\hat{\gamma}^{5}$, these
Lorentz transformations do not change the chirality of a wave function.

Now, note that Eq. (1) can be written as $(\mathrm{i}\hbar\hat{\gamma}^{0}\partial_{0}-\hat{\boldsymbol{\gamma}}\cdot\hat{\mathrm{\mathbf{p}}})\Psi=0$,
where $\hat{\mathrm{\mathbf{p}}}=-\mathrm{i}\hbar\hat{1}_{4}\nabla$
is the (Hermitian) Dirac momentum operator in (3+1) dimensions, and
$\hat{\boldsymbol{\gamma}}=(\hat{\gamma}^{1},\hat{\gamma}^{2},\hat{\gamma}^{3})$.
The latter equation can also be written as $(\mathrm{i}\hbar\hat{\gamma}^{5}\partial_{0}-\hat{\boldsymbol{\Sigma}}\cdot\hat{\mathrm{\mathbf{p}}})\Psi=0$,
where $\hat{\boldsymbol{\Sigma}}=\hat{\gamma}^{5}\hat{\gamma}^{0}\hat{\boldsymbol{\gamma}}=\hat{\boldsymbol{\Sigma}}^{\dagger}$.
Now, assuming that the operator $\mathrm{i}\hbar\hat{\gamma}^{5}\partial_{0}-\hat{\boldsymbol{\Sigma}}\cdot\hat{\mathrm{\mathbf{p}}}$
is acting on a typical plane-wave solution $\Psi_{\mathrm{\mathbf{p}}}$,
we obtain the algebraic relation $(\mathrm{sgn}(E)\left\Vert \mathrm{\mathbf{p}}\right\Vert \hat{\gamma}^{5}-\hat{\boldsymbol{\Sigma}}\cdot\mathrm{\mathbf{p}})\Psi_{\mathrm{\mathbf{p}}}=0$,
where $\mathrm{sgn}(E)$ is the sign of the relativistic energy of
the massless Dirac particle, i.e., $E=\mathrm{sgn}(E)\, c\left\Vert \mathrm{\mathbf{p}}\right\Vert $.
Thus, the (Hermitian) matrix $\hat{\boldsymbol{\Sigma}}\cdot\mathrm{\mathbf{p}}/\left\Vert \mathrm{\mathbf{p}}\right\Vert $
is related to $\hat{\gamma}^{5}$ via the aforementioned algebraic
relation. More generally, we could say that the (Hermitian) operator
$\hat{\lambda}\equiv\hat{\boldsymbol{\Sigma}}\cdot\hat{\mathrm{\mathbf{p}}}/\left\Vert \mathrm{\mathbf{p}}\right\Vert =\mathrm{diag}(\hat{\boldsymbol{\sigma}}\cdot\hat{\mathrm{\mathbf{p}}}_{[2]}/\left\Vert \mathrm{\mathbf{p}}_{[2]}\right\Vert ,\hat{\boldsymbol{\sigma}}\cdot\hat{\mathrm{\mathbf{p}}}_{[2]}/\left\Vert \mathrm{\mathbf{p}}_{[2]}\right\Vert )$
and the chirality matrix are linked (in the latter relation, $\mathrm{\mathbf{p}}_{[2]}$
indicates the eigenvalues of the operator $\hat{\mathrm{\mathbf{p}}}_{[2]}\equiv-\mathrm{i}\hbar\hat{1}_{2}\nabla$,
namely, the Weyl momentum operator, and $\left\Vert \mathrm{\mathbf{p}}_{[2]}\right\Vert =\left\Vert \mathrm{\mathbf{p}}\right\Vert $).
Finally, as is known, the matrix $\hat{\Sigma}^{j}/2$ is the infinitesimal
generator of rotation through an angle $\theta$ around the $x^{j}$-axis,
and this matrix is related to the generalization of the spin operator
in (3+1) dimensions, $\hat{\mathrm{S}}^{j}=\hbar\hat{\Sigma}^{j}/2$
(see appendix, subsection A). Thus, the operator $\hat{\lambda}$
pertains to the projection of the spin onto the direction of momentum
(which is not necessarily the direction of the particle motion) and
is called the helicity operator (see appendix, subsection B). 

Observe that, because $\mathrm{sgn}(E)\hat{\gamma}^{5}\Psi_{\mathrm{\mathbf{p}}}=\hat{\lambda}\Psi_{\mathrm{\mathbf{p}}}$,
$\Psi_{\mathrm{\mathbf{p}}}=(\Psi_{+})_{\mathrm{\mathbf{p}}}+(\Psi_{-})_{\mathrm{\mathbf{p}}}$
and $\hat{\gamma}^{5}(\Psi_{\pm})_{\mathrm{\mathbf{p}}}=(\pm1)(\Psi_{\pm})_{\mathrm{\mathbf{p}}}$,
we have $\hat{\lambda}(\Psi_{+})_{\mathrm{\mathbf{p}}}=\mathrm{sgn}(E)\hat{1}_{4}(\Psi_{+})_{\mathrm{\mathbf{p}}}$
and $\hat{\lambda}(\Psi_{-})_{\mathrm{\mathbf{p}}}=-\mathrm{sgn}(E)\hat{1}_{4}(\Psi_{-})_{\mathrm{\mathbf{p}}}$.
The eigenstate of $\hat{\lambda}$ with eigenvalue $+1$ is called
the right-handed (or right-helical) state (spin parallel to momentum),
and the eigenstate of $\hat{\lambda}$ with eigenvalue $-1$ is called
the left-handed (or left-helical) state (spin opposite to momentum)
\cite{RefP}. Clearly, for positive energies, right-handed and right-chiral,
as well as left-handed and left-chiral, can be considered as similar
concepts (naturally, within the present discussion), i.e., $(\Psi_{+})_{\mathrm{\mathbf{p}}}$
and $(\Psi_{-})_{\mathrm{\mathbf{p}}}$ are eigenstates of $\hat{\lambda}$
and $\hat{\gamma}^{5}$ with eigenvalues $+1$ and $-1$, respectively.
From the two eigenvalue equations for the operator $\hat{\lambda}$,
two eigenvalue equations for the operator $\hat{\lambda}_{[2]}\equiv\hat{\boldsymbol{\sigma}}\cdot\hat{\mathrm{\mathbf{p}}}_{[2]}/\left\Vert \mathrm{\mathbf{p}}_{[2]}\right\Vert $
are obtained, namely, $\hat{\lambda}_{[2]}(\varphi_{1})_{\mathrm{\mathbf{p}}}=\mathrm{sgn}(E)\hat{1}_{2}(\varphi_{1})_{\mathrm{\mathbf{p}}}$
and $\hat{\lambda}_{[2]}(\varphi_{2})_{\mathrm{\mathbf{p}}}=-\mathrm{sgn}(E)\hat{1}_{2}(\varphi_{2})_{\mathrm{\mathbf{p}}}$.
As expected, the latter two equations are precisely Eqs. (3) and (4)
(with the latter for plane-wave eigensolutions). Because $\hbar\hat{\boldsymbol{\sigma}}/2$
($\equiv\hat{\mathrm{\mathbf{S}}}_{[2]}$) is the spin operator for
two-component wave functions, the operator $\hat{\lambda}_{[2]}$
can be considered as the helicity operator for this type of wave function.
Thus, assuming or postulating that $(\varphi_{1})_{\mathrm{\mathbf{p}}}$
describes the antineutrino, we can say that the helicity of this positive-energy
particle is positive, i.e., the antineutrino is right-handed (the
latter fact has been determined experimentally). However, the same
equation for $(\varphi_{1})_{\mathrm{\mathbf{p}}}$ also tells us
that the helicity of the negative-energy antineutrino is negative,
and according to the so-called hole theory, this last result is interpreted
as the helicity of the positive-energy neutrino also being negative;
thus, the neutrino would be left-handed (the latter fact has also
been determined experimentally) \cite{RefI,RefJ}. Alternatively,
assuming that $(\varphi_{2})_{\mathrm{\mathbf{p}}}$ describes the
neutrino, we can say that the helicity of this positive-energy particle
is negative; thus, the conclusion is yet again that the neutrino is
left-handed. The same equation for $(\varphi_{2})_{\mathrm{\mathbf{p}}}$
also tells us that the helicity of the negative-energy neutrino is
positive, and according to hole theory, this last result is interpreted
as the helicity of the positive-energy antineutrino also being positive;
thus, the conclusion is yet again that the antineutrino is right-handed
\cite{RefI,RefJ}.

Finally, as is known, the Dirac equation (1) can provide real-valued
solutions as long as the Dirac gamma matrices satisfy $(\mathrm{i}\hat{\gamma}^{\mu})^{*}=\mathrm{i}\hat{\gamma}^{\mu}$;
this is precisely the condition that defines the Majorana representation
of the Dirac matrices (the asterisk $^{*}$ denotes a complex conjugate,
as usual). On the other hand, because the term $\mathrm{i}\hat{\sigma}^{0}$
in the Weyl equation (3) satisfies $(\mathrm{i}\hat{\sigma}^{0})^{*}=-\mathrm{i}\hat{\sigma}^{0}$,
this equation can give real-valued solutions only if the matrices
$\hat{\sigma}^{\mu}$ for $\mu=1,2,3$ in Eq. (3) also satisfy $(\mathrm{i}\hat{\sigma}^{\mu})^{*}=-\mathrm{i}\hat{\sigma}^{\mu}$,
in which case we obtain $\mu=1,3$ (the latter results for the matrices
$\hat{\sigma}^{\mu}$ are obviously also valid for the matrices $\hat{\bar{\sigma}}^{\mu}$
present in the Weyl equation (4)). Thus, although the (free) Weyl
equations are considered to describe massless neutral fermions, the
solutions of these equations are not always real. For example, the
solutions $\varphi_{1}=\varphi_{1}(y,t)$ ($\Rightarrow\partial_{1}\varphi_{1}=\partial_{3}\varphi_{1}=0$)
of Eq. (3), i.e., of the equation $(\mathrm{i}\hat{\sigma}^{0}\partial_{0}+\mathrm{i}\hat{\sigma}^{2}\partial_{2})\varphi_{1}=0$,
are always complex-valued. The same goes for the solutions $\varphi_{2}=\varphi_{2}(y,t)$
of Eq. (4) (in this case, Eq. (4) is $(\mathrm{i}\hat{\bar{\sigma}}^{0}\partial_{0}+\mathrm{i}\hat{\bar{\sigma}}^{2}\partial_{2})\varphi_{2}=0$).
Certainly, because the Weyl equations (Eqs. (3) and (4)) can explicitly
arise by using the Weyl representation in the Dirac equation (Eq.
(1)), it is expected that these equations will also generate complex
solutions. If the Majorana representation is used in the Dirac equation,
then a real system of coupled equations is obtained (which is why
it admits real-valued solutions), not two non-equivalent explicitly
covariant complex first-order equations \cite{RefN}.

\subsection{In (1+1) dimensions}

\noindent The equation for a single free massless Dirac particle in
(1+1) dimensions (or the one-dimensional free massless Dirac particle)
has the form
\begin{equation}
\mathrm{i}\hat{\gamma}^{\mu}\partial_{\mu}\Psi=0,
\end{equation}
where $\Psi$ is a two-component wave function, and the Dirac matrices
$\hat{\gamma}^{\mu}$, with $\mu=0,1$, satisfy the relations $\hat{\gamma}^{\mu}\hat{\gamma}^{\nu}+\hat{\gamma}^{\nu}\hat{\gamma}^{\mu}=2g^{\mu\nu}\hat{1}_{2}$,
where $g^{\mu\nu}=\mathrm{diag}(1,-1)$ ($\hat{1}_{2}$ is the $2\times2$
identity matrix), and $(\hat{\gamma}^{\mu})^{\dagger}=\hat{\gamma}^{0}\hat{\gamma}^{\mu}\hat{\gamma}^{0}$.
Here, we utilize the coordinates $x^{0}=ct$ and $x^{1}=x$ and $\partial_{\mu}=(c^{-1}\partial_{t},\partial_{x})$,
as usual. In the Weyl representation, the two-component wave function
and the Dirac matrices can be written as follows \cite{RefN}:
\begin{equation}
\Psi\equiv\left[\begin{array}{c}
\varphi_{1}\\
\varphi_{2}
\end{array}\right]\,,\quad\hat{\gamma}^{0}=\hat{\sigma}_{x}\,,\quad\hat{\gamma}^{1}=-\mathrm{i}\hat{\sigma}_{y}.
\end{equation}

By substituting $\Psi$, $\hat{\gamma}^{0}$ and $\hat{\gamma}^{1}$
from Eq. (10) into Eq. (9), we derive two one-component equations,
namely,
\begin{equation}
\mathrm{i}\left(\partial_{0}+\partial_{1}\right)\varphi_{1}=0\,,\quad\mathrm{and}\quad\mathrm{i}\left(\partial_{0}-\partial_{1}\right)\varphi_{2}=0.
\end{equation}
These two equations would be the (free) Weyl equations in (1+1) dimensions
\cite{RefQ}, and each of them would describe -- let us say -- massless
neutral one-dimensional ``fermions'' (i.e., 1D uncharged Weyl particles).
Additionally, the one-component wave functions $\varphi_{1}$ and
$\varphi_{2}$ are transformed in two different ways under the Lorentz
boost. In effect, let us write this Lorentz boost with speed $v$
along the $x$-axis in the following way: $\left[\, ct'\; x'\,\right]^{\mathrm{T}}=\hat{\Lambda}\left[\, ct\; x\,\right]^{\mathrm{T}}$
(i.e., $x^{\mu}\,'=\Lambda_{\;\,\nu}^{\mu}\, x^{\nu}$$\,\Rightarrow x^{\mu}=(\Lambda_{j}^{-1})_{\;\,\nu}^{\mu}\, x^{\nu}\,'$),
where $\hat{\Lambda}=\hat{\Lambda}(\omega)=\exp(\mathrm{i}\omega\hat{K})$,
with $\hat{K}=\mathrm{i}\hat{\sigma}_{x}$, $\tanh(\omega)=v/c\equiv\beta$
and $\cosh(\omega)=(1-\beta^{2})^{-1/2}\equiv\gamma$ ($\omega\in\mathbb{R}$).
Then, under this transformation, the Dirac wave function in (1+1)
dimensions transforms as $\Psi'(x',t')=\hat{S}(\Lambda)\Psi(x,t)$,
where $\hat{S}(\Lambda)=\exp(-\omega\hat{\gamma}^{0}\hat{\gamma}^{1}/2)=\exp(-\omega\hat{\sigma}_{z}/2)$,
and the latter matrix obeys the relation $\Lambda_{\;\,\nu}^{\mu}\hat{\gamma}^{\nu}=\hat{S}^{-1}(\Lambda)\hat{\gamma}^{\mu}\hat{S}(\Lambda)$.
Then, the matrix $\hat{S}(\Lambda)$ is a diagonal matrix (as expected),
and we obtain the following results:
\begin{equation}
\varphi_{1}'(x',t')=\left[\,\cosh\left(\frac{\omega}{2}\right)-\sinh\left(\frac{\omega}{2}\right)\right]\varphi_{1}(x,t)
\end{equation}
and 
\begin{equation}
\varphi_{2}'(x',t')=\left[\,\cosh\left(\frac{\omega}{2}\right)+\sinh\left(\frac{\omega}{2}\right)\right]\varphi_{2}(x,t).
\end{equation}
Thus, we have two different types of one-component Weyl wave function
in (1+1) dimensions. That is, the result in Eq. (12) is for one type
of 1D Weyl particle, and that of Eq. (13) is for the other type of
1D Weyl particle. On the other hand, by comparing the relations in
(7) and (8) with those in (12) and (13), we see that the two-component
wave function for a one-dimensional Dirac particle could transform
in a similar way to the two-component wave function for a specific
type of 3D Weyl particle. This is the case, for example, when the
one-dimensional Dirac particle is constrained to the  $z$-axis, in
which case its respective wave function, $\Psi(z,t)\equiv\left[\,\varphi_{1}(z,t)\;\varphi_{2}(z,t)\,\right]^{\mathrm{T}}$,
is exactly transformed as the Weyl wave function $\varphi_{1}(z,t)\equiv\left[\,\varphi_{1}^{\mathrm{t}}(z,t)\;\varphi_{1}^{\mathrm{b}}(z,t)\,\right]^{\mathrm{T}}$
(see Eq. (7)). The Dirac wave function $\Psi(z,t)$ would also transform
as the Weyl wave function $\varphi_{2}(z,t)\equiv\left[\,\varphi_{2}^{\mathrm{t}}(z,t)\;\varphi_{2}^{\mathrm{b}}(z,t)\,\right]^{\mathrm{T}}$
if the replacement $\omega\rightarrow-\omega$ is made in the function
given by Eq. (8) (i.e., if the relative speed between the Lorentz
frames changes from $v$ to $-v$). 

In (1+1) dimensions, the matrix $\hat{\Gamma}^{5}\equiv-\mathrm{i}\hat{\gamma}_{[2]}^{5}$,
where $\hat{\gamma}_{[2]}^{5}\equiv\mathrm{i}\hat{\gamma}^{0}\hat{\gamma}^{1}$,
is the chirality matrix, i.e., $\hat{\Gamma}^{5}=\hat{\sigma}_{z}$.
As we know, this matrix is Hermitian and satisfies the relations $(\hat{\Gamma}^{5})^{2}=\hat{1}_{2}$
and $\hat{\Gamma}^{5}\hat{\gamma}^{\mu}+\hat{\gamma}^{\mu}\hat{\Gamma}^{5}=\hat{0}_{2}$
\cite{RefN,RefR}. Thus, the ($2\times2$) chirality matrix in (1+1)
dimensions satisfies the same (three) basic properties as the ($4\times4$)
chirality matrix in (3+1) dimensions, as should be the case. Evidently,
the two-component Dirac wave functions $\Psi_{+}=\left[\,\varphi_{1}\;0\,\right]^{\mathrm{T}}=\frac{1}{2}(\hat{1}_{2}+\hat{\Gamma}^{5})\Psi$
(which must also satisfy the relations $\frac{1}{2}(\hat{1}_{2}+\hat{\Gamma}^{5})\Psi_{+}=\Psi_{+}$
and $\frac{1}{2}(\hat{1}_{2}-\hat{\Gamma}^{5})\Psi_{+}=0$) and $\Psi_{-}=\left[\,0\;\varphi_{2}\,\right]^{\mathrm{T}}=\frac{1}{2}(\hat{1}_{2}-\hat{\Gamma}^{5})\Psi$
(which must also satisfy the relations $\frac{1}{2}(\hat{1}_{2}-\hat{\Gamma}^{5})\Psi_{-}=\Psi_{-}$
and $\frac{1}{2}(\hat{1}_{2}+\hat{\Gamma}^{5})\Psi_{-}=0$) are eigenstates
of $\hat{\Gamma}^{5}$. Again, $\Psi_{+}$ is called the right-chiral
eigenstate (eigenvalue $+1$) and $\Psi_{-}$ the left-chiral eigenstate
(eigenvalue $-1$). Certainly, because the matrices $\hat{S}(\Lambda)$
and $\hat{\Gamma}^{5}$ commute, the Lorentz boost does not change
the chirality of the wave function.

Note that Eq. (9) can be written as $(\mathrm{i}\hbar\hat{\gamma}^{0}\partial_{0}-\hat{\gamma}^{1}\hat{\mathrm{p}})\Psi=0$,
where $\hat{\mathrm{p}}=-\mathrm{i}\hbar\hat{1}_{2}\partial_{1}$
is the (Hermitian) Dirac momentum operator in (1+1) dimensions. The
latter equation can also be written as $(\mathrm{i}\hbar\hat{\Gamma}^{5}\partial_{0}-\hat{\mathrm{p}})\Psi=0$
(remember that $\hat{\Gamma}^{5}=\hat{\gamma}^{0}\hat{\gamma}^{1}$).
Now, assuming that the operator $\mathrm{i}\hbar\hat{\Gamma}^{5}\partial_{0}-\hat{\mathrm{p}}$
is acting on a common plane-wave eigensolution $\Psi_{\mathrm{p}}$,
we obtain the algebraic relation $(\mathrm{sgn}(E)\left|\mathrm{p}\right|\hat{\Gamma}^{5}-\mathrm{p}\hat{1}_{2})\Psi_{\mathrm{p}}=0$,
where $\mathrm{sgn}(E)$ is the sign of the relativistic energy of
the one-dimensional massless Dirac particle, i.e., $E=\mathrm{sgn}(E)\, c\left|\mathrm{p}\right|$.
Thus, in (1+1) dimensions, the $2\times2$ matrix $\mathrm{p}\hat{1}_{2}/\left|\mathrm{p}\right|$
is related to $\hat{\Gamma}^{5}$ via the aforementioned algebraic
relation. We could say far more generally that there is a connection
between the operator $\hat{\mathrm{p}}/\left|\mathrm{p}\right|=\mathrm{diag}(\hat{\mathrm{p}}_{[1]}/\left|\mathrm{p}_{[1]}\right|,\hat{\mathrm{p}}_{[1]}/\left|\mathrm{p}_{[1]}\right|)$
and the chirality matrix (in the latter expression, $\mathrm{p}_{[1]}$
indicates the eigenvalues of the operator $\hat{\mathrm{p}}_{[1]}\equiv-\mathrm{i}\hbar\partial_{1}$,
namely, the momentum operator that acts on one-component wave functions,
and $\left|\mathrm{p}_{[1]}\right|=\left|\mathrm{p}\right|$). 

Note that, because $\mathrm{sgn}(E)\hat{\Gamma}^{5}\Psi_{\mathrm{p}}=(\hat{\mathrm{p}}/\left|\mathrm{p}\right|)\Psi_{\mathrm{p}}$,
$\Psi_{\mathrm{p}}=(\Psi_{+})_{\mathrm{p}}+(\Psi_{-})_{\mathrm{p}}$
and $\hat{\Gamma}^{5}(\Psi_{\pm})_{\mathrm{p}}=(\pm1)(\Psi_{\pm})_{\mathrm{p}}$,
we have that $(\hat{\mathrm{p}}/\left|\mathrm{p}\right|)(\Psi_{+})_{\mathrm{p}}=\mathrm{sgn}(E)\hat{1}_{2}(\Psi_{+})_{\mathrm{p}}$
and $(\hat{\mathrm{p}}/\left|\mathrm{p}\right|)(\Psi_{-})_{\mathrm{p}}=-\mathrm{sgn}(E)\hat{1}_{2}(\Psi_{-})_{\mathrm{p}}$.
Thus, in (1+1) dimensions, for positive energies, $(\Psi_{+})_{\mathrm{p}}$
and $(\Psi_{-})_{\mathrm{p}}$ are eigenstates of $\hat{\Gamma}^{5}$
and $\hat{\mathrm{p}}/\left|\mathrm{p}\right|$ with eigenvalues $+1$
and $-1$, respectively. From the latter two eigenvalue equations,
two other eigenvalue equations for the operator $\hat{\mathrm{p}}_{[1]}/\left|\mathrm{p}_{[1]}\right|$
are obtained, namely, $(\hat{\mathrm{p}}_{[1]}/\left|\mathrm{p}_{[1]}\right|)(\varphi_{1})_{\mathrm{p}}=\mathrm{sgn}(E)(\varphi_{1})_{\mathrm{p}}$
and $(\hat{\mathrm{p}}_{[1]}/\left|\mathrm{p}_{[1]}\right|)(\varphi_{2})_{\mathrm{p}}=-\mathrm{sgn}(E)(\varphi_{2})_{\mathrm{p}}$.
Certainly, the latter two equations are the equations in (11) for
the eigenstates of the energy and momentum. Thus, in (1+1) dimensions,
the momentum operator $\hat{\mathrm{p}}_{[1]}$ plays a role somewhat
similar to that of the helicity operator $\hat{\lambda}_{[2]}$ in
(3+1) dimensions (see appendix, subsection B). Naturally, in (1+1)
dimensions, there is no such thing as helicity or spin. Note that
the momentum of the positive-energy 1D Weyl particle described with
the wave function $(\varphi_{1})_{\mathrm{p}}$ is positive, and if
it has negative energy, then its momentum is negative. Likewise, the
momentum of the other positive-energy 1D Weyl particle, described
with the wave function $(\varphi_{2})_{\mathrm{p}}$, is negative,
and if it has negative energy, then its momentum is positive. In (1+1)
dimensions, the (Dirac) chiral plane-wave eigenstates $(\Psi_{+})_{\mathrm{p}}$
and $(\Psi_{-})_{\mathrm{p}}$ are such that the charge conjugate
of $(\Psi_{+})_{\mathrm{p}}$ ($(\Psi_{-})_{\mathrm{p}}$) is also
a right-chiral (left-chiral) state \cite{RefN}. Thus, the one-component
states $(\varphi_{1})_{\mathrm{p}}$ and $(\varphi_{2})_{\mathrm{p}}$
would not exactly describe a particle-antiparticle pair.

It should be noted, in passing, that unlike what happens with the
Weyl equations in (3+1) dimensions (Eqs. (3) and (4)), the Weyl equations
in (1+1) dimensions (equations in (11)) can always give real solutions,
i.e., the latter equations can have solutions \textit{\`a la} Majorana.
Certainly, complex solutions can also be obtained. 

Now, let us return to the Dirac equation in (9). Certainly, in choosing
a representation, one is choosing a set of Dirac matrices that satisfies
the Clifford relation. In addition to (\textit{i}), the Weyl (or chiral
or spinor) representation, $\{\hat{\gamma}^{0},\hat{\gamma}^{1}\}=\{\hat{\sigma}_{x},-\mathrm{i}\hat{\sigma}_{y}\}$
(with the Dirac wave function in Eq. (9) written as $\Psi\equiv\left[\,\varphi_{1}\;\varphi_{2}\,\right]^{\mathrm{T}}$),
we must consider three other representations in (1+1) dimensions,
namely, (\textit{ii}) the Dirac (or standard or Dirac-Pauli) representation,
$\{\hat{\gamma}^{0},\hat{\gamma}^{1}\}=\{\hat{\sigma}_{z},\mathrm{i}\hat{\sigma}_{y}\}$
(in this case, we write $\Psi\equiv\left[\,\varphi\;\chi\,\right]^{\mathrm{T}}$);
(\textit{iii}) the Majorana representation, $\{\hat{\gamma}^{0},\hat{\gamma}^{1}\}=\{\hat{\sigma}_{y},-\mathrm{i}\hat{\sigma}_{z}\}$
(in this case, $\Psi\equiv\left[\,\phi_{1}\;\phi_{2}\,\right]^{\mathrm{T}}$);
and (\textit{iv}) the Jackiw-Rebbi representation, $\{\hat{\gamma}^{0},\hat{\gamma}^{1}\}=\{\hat{\sigma}_{x},\mathrm{i}\hat{\sigma}_{z}\}$
(in this case, $\Psi\equiv\left[\,\chi_{1}\;\chi_{2}\,\right]^{\mathrm{T}}$)
\cite{RefS,RefT,RefU}. In the next section, we present some connections
between these representations, i.e., the connections between the Dirac
matrices and the two-component wave functions in two different representations.

\section{Boundary conditions for the 3D Weyl particle in a 1D box}

\noindent The Weyl equations (3) and (4), written compactly in Hamiltonian
form, are 
\begin{equation}
\mathrm{i}\hbar\hat{1}_{2}\frac{\partial}{\partial t}\varphi_{a}=\hat{\mathrm{H}}_{a}\varphi_{a}\,,\quad a=1,2,
\end{equation}
where 
\begin{equation}
\hat{\mathrm{H}}_{a}\equiv\sum_{j=1}^{j=3}\hat{\mathrm{H}}_{a,j}=-\mathrm{i}\hbar c(-1)^{a-1}\sum_{j=1}^{j=3}\hat{\sigma}_{j}\frac{\partial}{\partial x^{j}}=-\mathrm{i}\hbar c(-1)^{a-1}\,\hat{\boldsymbol{\sigma}}\cdot\nabla
\end{equation}
is the formally self-adjoint, or Hermitian, Weyl Hamiltonian operator
(or Dirac-Weyl operator), i.e., $\hat{\mathrm{H}}_{a}=\hat{\mathrm{H}}_{a}^{\dagger}$
($\dagger$ denotes the Hermitian conjugate, or the adjoint, of a
matrix and an operator). Naturally, this operator can also be written
in terms of the spin operator $\hat{\mathrm{\mathbf{S}}}_{[2]}$,
namely, $\hat{\mathrm{H}}_{a}=c\,\hat{\mathrm{\mathbf{S}}}_{[2]}\cdot\hat{\mathrm{\mathbf{p}}}_{[2]}/(-1)^{a-1}\frac{\hbar}{2}$
(as we know, the label $a$ indicates a particular type of 3D Weyl
particle). The latter expression for the operator $\hat{\mathrm{H}}_{a}$
could be generalized to the case of arbitrary spin (see Ref. \cite{RefD}
and references therein). In general, $\hat{\mathrm{H}}_{a}$ acts
on the two-component Weyl wave functions $\varphi_{a}=\varphi_{a}(\mathbf{\mathrm{\mathbf{r}}},t)$
that belong to the Hilbert space of square integrable functions, $\mathcal{H}=L^{2}(\Omega)^{2}$,
where $\Omega\subset\mathbb{R}^{3}$ represents a volume in three-dimensional
space. The scalar product in $\mathcal{H}$ is denoted by $\langle\psi_{a},\chi_{a}\rangle\equiv\int_{\Omega}\,\mathrm{d}^{3}\mathbf{\mathrm{\mathbf{r}}}\,\psi_{a}^{\dagger}\chi_{a}$,
and the norm is $\left\Vert \,\psi_{a}\,\right\Vert \equiv\sqrt{\langle\psi_{a},\psi_{a}\rangle}$,
as usual. The domain of the Hamiltonian, $\mathcal{D}(\hat{\mathrm{H}}_{a})$,
is the set of Weyl wave functions in $\mathcal{H}$ on which $\hat{\mathrm{H}}_{a}$
can act and (generally) includes the boundary conditions that these
wave functions must satisfy at the boundary of the volume $\Omega$.
Additionally, the Hamiltonian, when acting on one of these functions,
must produce a function that belongs to $\mathcal{H}$. By virtue
of one integration by parts, the self-adjointness condition (and therefore
the hermiticity condition) of $\hat{\mathrm{H}}_{a}$ is obtained,
namely, 
\begin{equation}
\langle\psi_{a},\hat{\mathrm{H}}_{a}\chi_{a}\rangle=\langle\hat{\mathrm{H}}_{a}\psi_{a},\chi_{a}\rangle-\mathrm{i}\hbar c(-1)^{a-1}\int_{\Omega}\,\mathrm{d}^{3}\mathbf{\mathrm{\mathbf{r}}}\,\,\nabla\cdot(\psi_{a}^{\dagger}\,\hat{\boldsymbol{\sigma}}\chi_{a})=\langle\hat{\mathrm{H}}_{a}\psi_{a},\chi_{a}\rangle,
\end{equation}
where the volume integral is the surface integral over the boundary
of the volume, i.e., $\oint_{\partial\Omega}\psi^{\dagger}\hat{\boldsymbol{\sigma}}\chi\cdot\mathrm{d}\boldsymbol{S}$
(because of the Gauss-Ostrogradsky theorem), and it must vanish because
one imposes some specific boundary conditions on $\psi_{a}$ and $\chi_{a}$
at $\partial\Omega$ that belong to $\mathcal{D}(\hat{\mathrm{H}}_{a})=\mathcal{D}(\hat{\mathrm{H}}_{a}^{\dagger})$.
Remember that, given $\hat{\mathrm{H}}_{a}$, the relation $\langle\psi_{a},\hat{\mathrm{H}}_{a}\chi_{a}\rangle=\langle\hat{\mathrm{H}}_{a}^{\dagger}\psi_{a},\chi_{a}\rangle$
is what essentially defines the adjoint operator $\hat{\mathrm{H}}_{a}^{\dagger}$
on a vector space. If $\hat{\mathrm{H}}_{a}$ is Hermitian (i.e.,
$\hat{\mathrm{H}}_{a}=\hat{\mathrm{H}}_{a}^{\dagger}$, and thus,
$\hat{\mathrm{H}}_{a}^{\dagger}$ acts in the same way as $\hat{\mathrm{H}}_{a}$),
then the relation $\langle\psi_{a},\hat{\mathrm{H}}_{a}\chi_{a}\rangle=\langle\hat{\mathrm{H}}_{a}\psi_{a},\chi_{a}\rangle$
is verified. If $\hat{\mathrm{H}}_{a}$ is self-adjoint in addition,
then it must be Hermitian, but also, the domains of $\hat{\mathrm{H}}_{a}$
and its corresponding adjoint must be equal (for example, if $\chi_{a}\in\mathcal{D}(\hat{\mathrm{H}}_{a})$
and $\psi_{a}\in\mathcal{D}(\hat{\mathrm{H}}_{a}^{\dagger})$, then
$\chi_{a}$ and $\psi_{a}$ must satisfy the same boundary condition). 

Now, let us consider the following three particular cases, i.e., the
following three particular Weyl Hamiltonian operators: 
\begin{equation}
\hat{\mathrm{H}}_{a,j}\equiv-\mathrm{i}\hbar c(-1)^{a-1}\,\hat{\sigma}_{j}\frac{\partial}{\partial x^{j}}\,,\quad j=1,2,3.
\end{equation}
Again, each of these operators is formally self-adjoint ($\hat{\mathrm{H}}_{a,j}=\hat{\mathrm{H}}_{a,j}^{\dagger}$)
and acts on the two-component wave functions $\varphi_{a,j}=\varphi_{a,j}(x^{j},t)$
that belong to $\mathcal{H}=L^{2}(\Omega)^{2}$, where $\Omega\subset\mathbb{R}^{3}$
is a three-dimensional square box. However, in each case, the corresponding
three-dimensional Weyl particle can only move inside an interval of
size $\ell$ (i.e., inside a one-dimensional box) on the $x^{j}$-axis,
with ends, for example, at $x^{j}=0$ and $x^{j}=\ell$ ($\Rightarrow\Omega_{j}=[0,\ell]$).
The scalar product in $\mathcal{H}$ becomes $\langle\psi_{a,j},\chi_{a,j}\rangle\equiv A_{j}\int_{\Omega_{j}}\,\mathrm{d}x^{j}\,\psi_{a,j}^{\dagger}\chi_{a,j}$,
where $A_{j}$ is the area of the side of the three-dimensional square
box perpendicular to the one-dimensional interval, as expected. In
this case, it can be demonstrated that the following relation is verified:
\begin{equation}
\langle\psi_{a,j},\hat{\mathrm{H}}_{a,j}\chi_{a,j}\rangle=\langle\hat{\mathrm{H}}_{a,j}\psi_{a,j},\chi_{a,j}\rangle-\left.\mathrm{i}\hbar c(-1)^{a-1}A_{j}\left[\,\psi_{a,j}^{\dagger}\hat{\sigma}_{j}\chi_{a,j}\,\right]\right|_{0}^{\ell},
\end{equation}
where $\left.\left[\, f\,\right]\right|_{0}^{\ell}\equiv f(x^{j}=\ell,t)-f(x^{j}=0,t)$.
Certainly, if the boundary conditions imposed on $\psi_{a,j}$ and
$\chi_{a,j}$ at the endpoints of the interval $\Omega_{j}$ lead
to cancellation of the boundary term in Eq. (18), then the operator
$\hat{\mathrm{H}}_{a,j}$ will be at least Hermitian. Precisely, the
most general family of self-adjoint boundary conditions for each of
the Weyl operators $\hat{\mathrm{H}}_{a,j}$ ($j=1,2,3$) is obtained
below from the most general families of self-adjoint boundary conditions
for three (free) Dirac operators in (1+1) dimensions.

Similarly, the Dirac equation (9) in Hamiltonian form is
\begin{equation}
\mathrm{i}\hbar\hat{1}_{2}\frac{\partial}{\partial t}\Psi=\hat{\mathrm{h}}\Psi,
\end{equation}
where 
\begin{equation}
\hat{\mathrm{h}}=-\mathrm{i}\hbar c\,\hat{\gamma}^{0}\hat{\gamma}^{1}\frac{\partial}{\partial x}
\end{equation}
is the formally self-adjoint, or Hermitian, (free) Dirac Hamiltonian
operator, $\hat{\mathrm{h}}=\hat{\mathrm{h}}^{\dagger}$. This operator
acts on the two-component Dirac wave functions $\Psi=\Psi(x,t)$ that
belong to the Hilbert space $\mathcal{H}=L^{2}([0,\ell])^{2}$. The
scalar product in $\mathcal{H}$ is denoted by $\langle\Phi,\chi\rangle\equiv\int_{[0,\ell]}\,\mathrm{d}x\,\Phi^{\dagger}\chi$,
and the norm is $\left\Vert \,\Phi\,\right\Vert \equiv\sqrt{\langle\Phi,\Phi\rangle}$,
as usual. In this case, it can be demonstrated that the following
relation is verified:
\begin{equation}
\langle\Phi,\hat{\mathrm{h}}\chi\rangle=\langle\hat{\mathrm{h}}\Phi,\chi\rangle-\left.\mathrm{i}\hbar c\left[\,\Phi^{\dagger}\hat{\gamma}^{0}\hat{\gamma}^{1}\chi\,\right]\right|_{0}^{\ell}.
\end{equation}
As we know, if the boundary conditions imposed on $\Phi$ and $\chi$
at the endpoints of the interval $[0,\ell]$ lead to cancellation
of the boundary term in Eq. (21), then the operator $\hat{\mathrm{h}}$
will be at least Hermitian. However, given a set of boundary conditions
imposed on $\chi\in\mathcal{D}(\hat{\mathrm{h}})$, if the cancellation
of the boundary term in Eq. (21) only depends on imposing the same
boundary conditions on $\Phi\in\mathcal{D}(\hat{\mathrm{h}}^{\dagger})$,
then the Hamiltonian is also a self-adjoint operator. 

Now, with the operator $\hat{\mathrm{h}}$ in mind, let us construct
the following three Dirac Hamiltonian operators:
\begin{equation}
\hat{\mathrm{h}}_{j}\equiv-\mathrm{i}\hbar c\,\hat{\sigma}_{j}\frac{\partial}{\partial x^{j}}\,,\quad j=1,2,3.
\end{equation}
Clearly, these three operators are essentially the three Weyl Hamiltonian
operators $\hat{\mathrm{H}}_{a,j}$ ($j=1,2,3$) in Eq. (17). Additionally,
note that $\hat{\mathrm{h}}_{j}$ satisfies the relation given in
Eq. (21) with the replacements $\hat{\mathrm{h}}\rightarrow\hat{\mathrm{h}}_{j}$,
$\hat{\gamma}^{0}\hat{\gamma}^{1}\rightarrow\hat{\sigma}_{j}$, $\Phi\rightarrow\Phi_{j}$
and $\chi\rightarrow\chi_{j}$, and the latter relation is similar
to the one that $\hat{\mathrm{H}}_{a,j}$ must satisfy (i.e., Eq.
(18)). On the other hand, note that $\hat{\mathrm{h}}_{1}$ is the
Dirac Hamiltonian for a one-dimensional Dirac particle in the interval
$x\in[0,\ell]$ in both the Dirac and Majorana representations ($\hat{\gamma}^{0}\hat{\gamma}^{1}=\hat{\sigma}_{x}$).
Likewise, $\hat{\mathrm{h}}_{2}$ is the Dirac Hamiltonian in the
interval $y\in[0,\ell]$, in the Jackiw-Rebbi representation ($\hat{\gamma}^{0}\hat{\gamma}^{1}=\hat{\sigma}_{y}$),
and $\hat{\mathrm{h}}_{3}$ is the Dirac Hamiltonian in the interval
$z\in[0,\ell]$ in the Weyl representation ($\hat{\gamma}^{0}\hat{\gamma}^{1}=\hat{\sigma}_{z}$).
Precisely, we already know which is the most general family of boundary
conditions for the self-adjoint operator $\hat{\mathrm{h}}_{3}$ (see
Refs. \cite{RefV,RefW} and references therein), namely, 
\begin{equation}
\left[\begin{array}{c}
\varphi_{1}(z=\ell,t)\\
\varphi_{2}(z=0,t)
\end{array}\right]=\hat{U}_{3}\left[\begin{array}{c}
\varphi_{2}(z=\ell,t)\\
\varphi_{1}(z=0,t)
\end{array}\right],
\end{equation}
where $\hat{U}_{3}$ is a unitary matrix; thus, this set of boundary
conditions is characterized by four real parameters. Certainly, the
latter result can be obtained using the von Neumann theory of self-adjoint
extensions of symmetric operators, although the construction of the
result can also be done using less rigorous arguments \cite{RefV}.

From the result in Eq. (23), we can find the most general set of boundary
conditions for the self-adjoint operator $\hat{\mathrm{h}}_{1}$,
which is written in both the Dirac and Majorana representations. For
example, the Hamiltonian $\hat{\mathrm{h}}_{1}$ in the Dirac representation
and $\hat{\mathrm{h}}_{3}$ (in the Weyl representation) are related
via a unitary transformation, namely, $\hat{S}_{13}$; as a consequence,
$\hat{\mathrm{h}}_{3}=\hat{S}_{13}\,\hat{\mathrm{h}}_{1}\,\hat{S}_{13}^{\dagger}$
(certainly, we are assuming that the spatial variable is the same),
and 
\begin{equation}
\left[\begin{array}{c}
\varphi_{1}\\
\varphi_{2}
\end{array}\right]=\hat{S}_{13}\left[\begin{array}{c}
\varphi\\
\chi
\end{array}\right]\,,\quad\mathrm{with}\quad\hat{S}_{13}=\frac{1}{\sqrt{2}}(\hat{\sigma}_{x}+\hat{\sigma}_{z})
\end{equation}
(see Ref. \cite{RefN}). The first relation in Eq. (24) allows us
to obtain the Dirac wave function associated with the operator $\hat{\mathrm{h}}_{3}$
from the Dirac wave function associated with the operator $\hat{\mathrm{h}}_{1}$
in the Dirac representation. Thus, the most general family of boundary
conditions for the self-adjoint operator $\hat{\mathrm{h}}_{1}$ is
obtained by substituting the entire result in Eq. (24) into Eq. (23)
and finally making the obvious replacements $z\rightarrow x$ and
$\hat{U}_{3}\rightarrow\hat{U}_{1}$, namely, 
\begin{equation}
\left[\begin{array}{c}
\varphi(x=\ell,t)+\chi(x=\ell,t)\\
\varphi(x=0,t)-\chi(x=0,t)
\end{array}\right]=\hat{U}_{1}\left[\begin{array}{c}
\varphi(x=\ell,t)-\chi(x=\ell,t)\\
\varphi(x=0,t)+\chi(x=0,t)
\end{array}\right],
\end{equation}
where $\hat{U}_{1}$ is the unitary matrix in this case. Certainly,
in the Majorana representation, the most general family of boundary
conditions for the self-adjoint operator $\hat{\mathrm{h}}_{1}$ has
the same format as that given in Eq. (25), namely,
\begin{equation}
\left[\begin{array}{c}
\phi_{1}(x=\ell,t)+\phi_{2}(x=\ell,t)\\
\phi_{1}(x=0,t)-\phi_{2}(x=0,t)
\end{array}\right]=\hat{U}'_{1}\left[\begin{array}{c}
\phi_{1}(x=\ell,t)-\phi_{2}(x=\ell,t)\\
\phi_{1}(x=0,t)+\phi_{2}(x=0,t)
\end{array}\right],
\end{equation}
where $\hat{U}'_{1}$ is also a unitary matrix. To demonstrate this,
we can follow the following simple procedure (which is identical  to
the one that led to the result given in Eq. (25)). Note, first, that
the Hamiltonian $\hat{\mathrm{h}}_{1}$ in the Majorana representation
and $\hat{\mathrm{h}}_{3}$ (in the Weyl representation) are also
related via a unitary transformation, namely, $\hat{S}'_{13}$; therefore,
$\hat{\mathrm{h}}_{3}=\hat{S}'_{13}\,\hat{\mathrm{h}}_{1}\,(\hat{S}'_{13})^{\dagger}$,
and  
\begin{equation}
\left[\begin{array}{c}
\varphi_{1}\\
\varphi_{2}
\end{array}\right]=\hat{S}'_{13}\left[\begin{array}{c}
\phi_{1}\\
\phi_{2}
\end{array}\right]\,,\quad\mathrm{with}\quad\hat{S}'_{13}=\frac{1}{2}\left(-\mathrm{i}\hat{1}_{2}+\hat{\sigma}_{x}+\hat{\sigma}_{y}+\hat{\sigma}_{z}\right)=\exp\left(-\mathrm{i}\frac{\pi}{4}\right)\frac{1}{\sqrt{2}}\left[\begin{array}{cc}
1 & 1\\
\mathrm{i} & -\mathrm{i}
\end{array}\right]
\end{equation}
(see Ref. \cite{RefN}). The first relation in Eq. (27) allows us
to obtain the Dirac wave function associated with the operator $\hat{\mathrm{h}}_{3}$
from the Dirac wave function associated with the operator $\hat{\mathrm{h}}_{1}$
in the Majorana representation. Then, the most general family of boundary
conditions for the self-adjoint operator $\hat{\mathrm{h}}_{1}$,
i.e., Eq. (26), is obtained by substituting the entire result in Eq.
(27) into Eq. (23), also making the obvious replacements $z\rightarrow x$
and $\hat{U}_{3}\rightarrow\hat{U}'_{1}$ and, finally, certain simplifications.

Similarly, from the result in Eq. (23), we can find the most general
set of boundary conditions for the self-adjoint operator $\hat{\mathrm{h}}_{2}$,
which is written in the Jackiw-Rebbi representation. In effect, the
Hamiltonians $\hat{\mathrm{h}}_{2}$ and $\hat{\mathrm{h}}_{3}$ (in
the Weyl representation) are related via the unitary matrix $\hat{S}_{23}$;
thus, $\hat{\mathrm{h}}_{3}=\hat{S}_{23}\,\hat{\mathrm{h}}_{2}\,\hat{S}_{23}^{\dagger}$,
and
\begin{equation}
\left[\begin{array}{c}
\varphi_{1}\\
\varphi_{2}
\end{array}\right]=\hat{S}_{23}\left[\begin{array}{c}
\chi_{1}\\
\chi_{2}
\end{array}\right]\,,\quad\mathrm{with}\quad\hat{S}_{23}=\frac{1}{\sqrt{2}}(\hat{1}_{2}-\mathrm{i}\hat{\sigma}_{x}).
\end{equation}
The first relation in Eq. (28) allows us to obtain the Dirac wave
function associated with the operator $\hat{\mathrm{h}}_{3}$ from
the Dirac wave function associated with the operator $\hat{\mathrm{h}}_{2}$.
Thus, the most general family of boundary conditions for the self-adjoint
operator $\hat{\mathrm{h}}_{2}$ is obtained by substituting the entire
result in Eq. (28) into Eq. (23) and finally making the obvious replacements
$z\rightarrow y$ and $\hat{U}_{3}\rightarrow\hat{U}_{2}$, namely,
\begin{equation}
\left[\begin{array}{c}
\chi_{1}(y=\ell,t)-\mathrm{i}\chi_{2}(y=\ell,t)\\
\chi_{2}(y=0,t)-\mathrm{i}\chi_{1}(y=0,t)
\end{array}\right]=\hat{U}_{2}\left[\begin{array}{c}
\chi_{2}(y=\ell,t)-\mathrm{i}\chi_{1}(y=\ell,t)\\
\chi_{1}(y=0,t)-\mathrm{i}\chi_{2}(y=0,t)
\end{array}\right],
\end{equation}
where $\hat{U}_{2}$ is a unitary matrix.

From the results given in Eqs. (23), (25) (or (26)) and (29), we can
immediately write the most general families of boundary conditions
for the self-adjoint Weyl operators $\hat{\mathrm{H}}_{a,3}$, $\hat{\mathrm{H}}_{a,1}$
and $\hat{\mathrm{H}}_{a,2}$, respectively. First, the operator 
\begin{equation}
\hat{\mathrm{H}}_{a,1}\equiv-\mathrm{i}\hbar c(-1)^{a-1}\,\hat{\sigma}_{x}\frac{\partial}{\partial x}\,\quad(a=1,2)
\end{equation}
can act on two-component wave functions $\varphi_{a,1}\equiv\varphi_{a}=\left[\,\varphi_{a}^{\mathrm{t}}\;\varphi_{a}^{\mathrm{b}}\,\right]^{\mathrm{T}}$
that satisfy any of the following boundary conditions: 
\begin{equation}
\left[\begin{array}{c}
\varphi_{a}^{\mathrm{t}}(x=\ell,t)+\varphi_{a}^{\mathrm{b}}(x=\ell,t)\\
\varphi_{a}^{\mathrm{t}}(x=0,t)-\varphi_{a}^{\mathrm{b}}(x=0,t)
\end{array}\right]=\hat{A}_{1}\left[\begin{array}{c}
\varphi_{a}^{\mathrm{t}}(x=\ell,t)-\varphi_{a}^{\mathrm{b}}(x=\ell,t)\\
\varphi_{a}^{\mathrm{t}}(x=0,t)+\varphi_{a}^{\mathrm{b}}(x=0,t)
\end{array}\right],
\end{equation}
where $\hat{A}_{1}$ is a unitary matrix. Because we want to highlight
here the dependence of the boundary conditions with the label $a$,
we can assume that the operator 
\begin{equation}
\hat{\mathrm{H}}_{a,2}\equiv-\mathrm{i}\hbar c(-1)^{a-1}\,\hat{\sigma}_{y}\frac{\partial}{\partial y}\,\quad(a=1,2)
\end{equation}
can act on two-component wave functions that we write again as  $\varphi_{a,2}\equiv\varphi_{a}=\left[\,\varphi_{a}^{\mathrm{t}}\;\varphi_{a}^{\mathrm{b}}\,\right]^{\mathrm{T}}$
and that satisfy any of the following boundary conditions:
\begin{equation}
\left[\begin{array}{c}
\varphi_{a}^{\mathrm{t}}(y=\ell,t)-\mathrm{i}\varphi_{a}^{\mathrm{b}}(y=\ell,t)\\
\varphi_{a}^{\mathrm{b}}(y=0,t)-\mathrm{i}\varphi_{a}^{\mathrm{t}}(y=0,t)
\end{array}\right]=\hat{A}_{2}\left[\begin{array}{c}
\varphi_{a}^{\mathrm{b}}(y=\ell,t)-\mathrm{i}\varphi_{a}^{\mathrm{t}}(y=\ell,t)\\
\varphi_{a}^{\mathrm{t}}(y=0,t)-\mathrm{i}\varphi_{a}^{\mathrm{b}}(y=0,t)
\end{array}\right],
\end{equation}
where $\hat{A}_{2}$ is a unitary matrix. Similarly, the operator
\begin{equation}
\hat{\mathrm{H}}_{a,3}\equiv-\mathrm{i}\hbar c(-1)^{a-1}\,\hat{\sigma}_{z}\frac{\partial}{\partial z}\,\quad(a=1,2)
\end{equation}
acts on the wave functions $\varphi_{a,3}\equiv\varphi_{a}=\left[\,\varphi_{a}^{\mathrm{t}}\;\varphi_{a}^{\mathrm{b}}\,\right]^{\mathrm{T}}$
that satisfy at least one of  the following infinite boundary conditions:
\begin{equation}
\left[\begin{array}{c}
\varphi_{a}^{\mathrm{t}}(z=\ell,t)\\
\varphi_{a}^{\mathrm{b}}(z=0,t)
\end{array}\right]=\hat{A}_{3}\left[\begin{array}{c}
\varphi_{a}^{\mathrm{b}}(z=\ell,t)\\
\varphi_{a}^{\mathrm{t}}(z=0,t)
\end{array}\right],
\end{equation}
where $\hat{A}_{3}$ is a unitary matrix. 

Note that the Weyl equations with the operators $\hat{\mathrm{H}}_{a,1}$
and $\hat{\mathrm{H}}_{a,3}$, i.e., Eq. (14) with the replacements
$\hat{\mathrm{H}}_{a}\rightarrow\hat{\mathrm{H}}_{a,1}$ and $\hat{\mathrm{H}}_{a}\rightarrow\hat{\mathrm{H}}_{a,3}$,
can provide real solutions (see the comment made in the last paragraph
of section II, subsection A). Thus, if we impose on the corresponding
wave function $\varphi_{a}$ the reality condition ($\varphi_{a}=\varphi_{a}^{*}$),
$\varphi_{a}$ and $\varphi_{a}^{*}$ must meet the boundary conditions
in Eqs. (31) and (35), in which case the unitary matrices in these
equations, $\hat{A}_{1}$ and $\hat{A}_{3}$, must also be real, i.e.,
these matrices must be orthogonal. Consequently, in this case, the
families of general boundary conditions given in Eqs. (31) and (35)
only depend on one real parameter. On the other hand, the Weyl equations
in (14) with the replacement $\hat{\mathrm{H}}_{a}\rightarrow\hat{\mathrm{H}}_{a,2}$
cannot give real-valued solutions. Thus, these necessarily complex
solutions support any boundary condition included in the real four-parameter
general family of boundary conditions given in Eq. (33) (see appendix,
subsection C).

\section{Boundary conditions for the 1D Weyl particle in a (1D) box}

\noindent We have described above general families of self-adjoint
boundary conditions for a 3D Weyl particle in three one-dimensional
boxes. Let us now consider a one-dimensional Weyl particle in a box
of size $\ell$, with ends, for example, at $x=0$ and $x=\ell$.
First, the two Weyl equations in Eq. (11) can be written in a single
equation in their canonical form as follows:
\begin{equation}
\mathrm{i}\hbar\frac{\partial}{\partial t}\varphi_{a}=\hat{\mathrm{h}}_{a}\varphi_{a}\,,\quad a=1,2,
\end{equation}
where
\begin{equation}
\hat{\mathrm{h}}_{a}\equiv-\mathrm{i}\hbar c(-1)^{a-1}\frac{\partial}{\partial x}
\end{equation}
is the formally self-adjoint, or Hermitian, one-dimensional Weyl Hamiltonian
operator, i.e., $\hat{\mathrm{h}}_{a}=\hat{\mathrm{h}}_{a}^{\dagger}$.
Note that $\hat{\mathrm{h}}_{a}$ is practically the non-relativistic
momentum operator \cite{RefX}, i.e., $\hat{\mathrm{h}}_{a}=c(-1)^{a-1}\hat{\mathrm{p}}_{[1]}$
(as we know, the label $a$ indicates the type of 1D Weyl particle
that we are describing). The Hamiltonian is also a self-adjoint operator;
this is (essentially) because its domain, i.e., the set of Weyl one-component
wave functions $\varphi_{a}=\varphi_{a}(x,t)$ in the Hilbert space
of the square integrable functions $\mathcal{H}=L^{2}[0,\ell]$ on
which $\hat{\mathrm{h}}_{a}$ can act ($\equiv\mathcal{D}(\hat{\mathrm{h}}_{a})\subset\mathcal{H}$),
includes the generalized periodic boundary condition dependent on
a single parameter, namely,
\begin{equation}
\varphi_{a}(\ell,t)=\exp(\mathrm{i}\eta)\,\varphi_{a}(0,t),
\end{equation}
with $\eta\in[0,2\pi)$; in addition, $\hat{\mathrm{h}}_{a}\varphi_{a}\in\mathcal{H}$
(for a complete derivation of the latter result, see Ref. \cite{RefX}).
Moreover, the scalar product in $\mathcal{H}$ is denoted by $\langle\psi_{a},\chi_{a}\rangle\equiv\int_{0}^{\ell}\,\mathrm{d}x\,\psi_{a}^{*}\chi_{a}$,
and the norm is $\left\Vert \,\psi_{a}\,\right\Vert \equiv\sqrt{\langle\psi_{a},\psi_{a}\rangle}$.
Precisely, $\hat{\mathrm{h}}_{a}$ satisfies the following condition,
that is, the hermiticity condition (or, in this case, the self-adjointness
condition):
\begin{equation}
\langle\psi_{a},\hat{\mathrm{h}}_{a}\chi_{a}\rangle=\langle\hat{\mathrm{h}}_{a}\psi_{a},\chi_{a}\rangle-\left.\mathrm{i}\hbar c(-1)^{a-1}\left[\,\psi_{a}^{*}\chi_{a}\,\right]\right|_{0}^{\ell}=\langle\hat{\mathrm{h}}_{a}\psi_{a},\chi_{a}\rangle,
\end{equation}
where $\psi_{a}$ and $\chi_{a}$ are Weyl wave functions in $\mathcal{D}(\hat{\mathrm{h}}_{a})=\mathcal{D}(\hat{\mathrm{h}}_{a}^{\dagger})$. 

Note that the (free) Weyl equations in (1+1) dimensions (see Eq. (36))
can always provide real-valued solutions. Thus, if we impose on the
wave function $\varphi_{a}$ the reality condition, i.e., $\varphi_{a}=\varphi_{a}^{*}$,
then $\varphi_{a}$ and $\varphi_{a}^{*}$ must satisfy the same boundary
condition written above, in which case the phase factor $\exp(\mathrm{i}\eta)$
must be real. The latter condition implies that $\eta=0$ or $\eta=\pi$.
Thus, the boundary conditions for the (uncharged) 1D (free) Weyl particle
are $\varphi_{a}(\ell,t)=\varphi_{a}(0,t)$, i.e., the periodic boundary
condition, and $\varphi_{a}(\ell,t)=-\varphi_{a}(0,t)$, i.e., the
antiperiodic boundary condition. 

\section{Conclusions}

\noindent Our primary objective in this article was to obtain the
most general families of (self-adjoint) boundary conditions that can
be imposed on the general solutions of the time-dependent Weyl equations
that describe a 3D Weyl particle and a 1D Weyl particle in a one-dimensional
box. Because the one-dimensional box can be placed on any Cartesian
axis, one has three Weyl equations for the 3D Weyl particle in the
box (i.e., Eq. (14) with the replacement $\hat{\mathrm{H}}_{a}\rightarrow\hat{\mathrm{H}}_{a,j}$,
where $\hat{\mathrm{H}}_{a,j}$ is given in Eq. (17)). Each of the
Weyl Hamiltonians present in these equations can be identified with
a Dirac Hamiltonian that describes a 1D Dirac particle in a one-dimensional
box. Because we know which is the most general family of boundary
conditions for the Dirac operator in the Weyl representation, we were
able to construct, from the latter, the most general families of boundary
conditions for the other two Dirac operators by means of unitary transformations
(i.e., via changes of representation). In the end, the three general
families of (self-adjoint) boundary conditions for the 3D Weyl particle
in the 1D box are characterized by four real parameters (which, in
each case, constitute a $2\times2$ unitary matrix). In the cases
where Weyl's equations can give real-valued general solutions, each
family of four real parameters becomes two families each characterized
by a single real parameter (each parameter within a $2\times2$ orthogonal
matrix).

On the other hand, for the 1D Weyl particle, we have a Weyl equation
whose Weyl Hamiltonian is similar to the non-relativistic momentum
operator for the particle in a box (see Eqs. (36) and (37)). Thus,
the most general family of (self-adjoint) boundary conditions for
the 1D Weyl particle in the (1D) box is characterized by one real
parameter. In this case, the general solutions of the Weyl equation
can always be real-valued, but then these solutions can only accept
periodic and antiperiodic boundary conditions.

The mathematical manner in which the Weyl Hamiltonian (and that of
Dirac) acts on a wave function in (3+1) dimensions, that is, in a
way that depends on the direction in which the particles move and
on certain matrices along that direction (the ``$\hat{\boldsymbol{\sigma}}\cdot\hat{\mathrm{\mathbf{p}}}_{[2]}$''
term in the Weyl Hamiltonian), has found applications that far surpass
high-energy physics. For example, in condensed-matter physics, the
Weyl and the Dirac equations can be used to describe the band structure
of Dirac materials (or systems where the low-energy electronic excitations
are essentially described by Weyl or Dirac equations). A famous two-dimensional
Dirac material is graphene, which hosts excitations described by a
2D Weyl equation (see, for example, Refs. \cite{RefY,RefZ} and references
therein). We hope that our results, although valid for a one-dimensional
system, can also be applied to some of the various accessible systems
within condensed-matter physics.

\section{Appendix}

\subsection{On rotations }

\noindent Let us write an ordinary spatial rotation through an angle
$\theta$ around the $x^{j}$-axis, that is, $\left[\, ct'\; x'\; y'\; z'\,\right]^{\mathrm{T}}=\hat{\Lambda}_{j}\left[\, ct\; x\; y\; z\,\right]^{\mathrm{T}}$
(i.e., $x^{\mu}\,'=(\Lambda_{j})_{\;\,\nu}^{\mu}\, x^{\nu}\Rightarrow x^{\mu}=(\Lambda_{j}^{-1})_{\;\,\nu}^{\mu}\, x^{\nu}\,'$),
where
\[
\hat{\Lambda}_{j}=\hat{\Lambda}_{j}(\theta)=\exp\left(\mathrm{i}\theta\hat{J}_{j}\right),\tag{A1}
\]
with
\[
\hat{J}_{1}=\left[\begin{array}{cc}
\hat{0}_{2} & \hat{0}_{2}\\
\hat{0}_{2} & \hat{\sigma}_{y}
\end{array}\right]\,,\quad\hat{J}_{2}=\left[\begin{array}{cc}
\hat{0}_{2} & \frac{\mathrm{i}}{2}(\hat{1}_{2}-\hat{\sigma}_{z})\\
-\frac{\mathrm{i}}{2}(\hat{1}_{2}-\hat{\sigma}_{z}) & \hat{0}_{2}
\end{array}\right]\,,\quad\hat{J}_{3}=\left[\begin{array}{cc}
\hat{0}_{2} & -\frac{\mathrm{i}}{2}(\hat{\sigma}_{x}-\mathrm{i}\hat{\sigma}_{y})\\
\frac{\mathrm{i}}{2}(\hat{\sigma}_{x}+\mathrm{i}\hat{\sigma}_{y}) & \hat{0}_{2}
\end{array}\right].\tag{A2}
\]
Then, under this linear transformation, the Dirac wave function in
(3+1) dimensions transforms as $\Psi'(x^{k}\,',t')=\hat{S}(\Lambda_{j})\Psi(x^{k},t)$
($k=1,2,3$), where $\hat{S}(\Lambda_{j})$, which obeys the relation
$(\Lambda_{j})_{\;\,\nu}^{\mu}\hat{\gamma}^{\nu}=\hat{S}^{-1}(\Lambda_{j})\,\hat{\gamma}^{\mu}\,\hat{S}(\Lambda_{j})$,
is given by $\hat{S}(\Lambda_{j})=\exp(\mathrm{i}\theta\,\mathrm{i}\hat{\gamma}^{k}\hat{\gamma}^{l}/2)$
($l=1,2,3$) for cyclic $\{j,k,l\}$. Because $\hat{\boldsymbol{\Sigma}}=(\mathrm{i}\hat{\gamma}^{2}\hat{\gamma}^{3},\mathrm{i}\hat{\gamma}^{3}\hat{\gamma}^{1},\mathrm{i}\hat{\gamma}^{1}\hat{\gamma}^{2})$
($=\hat{\gamma}^{5}\hat{\gamma}^{0}\hat{\boldsymbol{\gamma}}$), we
can write the $4\times4$ matrix $\hat{S}(\Lambda_{j})$ as follows:
\[
\hat{S}(\Lambda_{j})=\exp\left(\mathrm{i}\theta\,\frac{\hat{\Sigma}^{j}}{2}\right),\tag{A3}
\]
i.e., the spin operator in (3+1) dimensions $\hat{\mathrm{\mathbf{S}}}=\hbar\hat{\boldsymbol{\Sigma}}/2$
is essentially the generator of spatial rotations. In the Weyl representation,
$\hat{S}(\Lambda_{j})$ is a block-diagonal matrix because $\hat{\boldsymbol{\Sigma}}=\mathrm{diag}(\hat{\boldsymbol{\sigma}},\hat{\boldsymbol{\sigma}})$,
and we obtain the following results:
\[
\left[\begin{array}{c}
\varphi_{1}^{\mathrm{t}}\,'(x^{j}\,',t')\\
\varphi_{1}^{\mathrm{b}}\,'(x^{j}\,',t')
\end{array}\right]=\left[\,\cos\left(\frac{\theta}{2}\right)\hat{1}_{2}+\mathrm{i}\sin\left(\frac{\theta}{2}\right)\hat{\sigma}_{j}\,\right]\left[\begin{array}{c}
\varphi_{1}^{\mathrm{t}}(x^{j},t)\\
\varphi_{1}^{\mathrm{b}}(x^{j},t)
\end{array}\right]\tag{A4}
\]
and 
\[
\left[\begin{array}{c}
\varphi_{2}^{\mathrm{t}}\,'(x^{j}\,',t')\\
\varphi_{2}^{\mathrm{b}}\,'(x^{j}\,',t')
\end{array}\right]=\left[\,\cos\left(\frac{\theta}{2}\right)\hat{1}_{2}+\mathrm{i}\sin\left(\frac{\theta}{2}\right)\hat{\sigma}_{j}\,\right]\left[\begin{array}{c}
\varphi_{2}^{\mathrm{t}}(x^{j},t)\\
\varphi_{2}^{\mathrm{b}}(x^{j},t)
\end{array}\right].
\]
That is, the two-component wave functions (or Weyl spinors) in (3+1)
dimensions $\varphi_{1}$ and $\varphi_{2}$ transform in the same
way under spatial rotations. Obviously, in (1+1) dimensions, a pure
(spatial) rotation is not possible.

\subsection{On the concept of helicity}

\noindent In (3+1) dimensions, the eigenstates $(\Psi_{+})_{\mathrm{\mathbf{p}}}=\left[\,(\varphi_{1})_{\mathrm{\mathbf{p}}}\;\,0\,\right]^{\mathrm{T}}$
and $(\Psi_{-})_{\mathrm{\mathbf{p}}}=\left[\,0\;\,(\varphi_{2})_{\mathrm{\mathbf{p}}}\,\right]^{\mathrm{T}}$
of the helicity operator $\hat{\lambda}\equiv\hat{\boldsymbol{\Sigma}}\cdot\hat{\mathrm{\mathbf{p}}}/\left\Vert \mathrm{\mathbf{p}}\right\Vert =\hat{\mathrm{\mathbf{S}}}\cdot\hat{\mathrm{\mathbf{p}}}/\frac{\hbar}{2}\left\Vert \mathrm{\mathbf{p}}\right\Vert $
($=\mathrm{diag}(\hat{\lambda}_{[2]},\hat{\lambda}_{[2]})$) satisfy
the following relations that depend on the sign of the energy: 
\[
\hat{\mathrm{\mathbf{S}}}\cdot\frac{\hat{\mathrm{\mathbf{p}}}}{\left\Vert \mathrm{\mathbf{p}}\right\Vert }\,(\Psi_{+})_{\mathrm{\mathbf{p}}}=\mathrm{sgn}(E)\,\frac{\hbar}{2}\,\hat{1}_{4}(\Psi_{+})_{\mathrm{\mathbf{p}}}\,,\quad\hat{\mathrm{\mathbf{S}}}\cdot\frac{\hat{\mathrm{\mathbf{p}}}}{\left\Vert \mathrm{\mathbf{p}}\right\Vert }\,(\Psi_{-})_{\mathrm{\mathbf{p}}}=-\mathrm{sgn}(E)\,\frac{\hbar}{2}\,\hat{1}_{4}(\Psi_{-})_{\mathrm{\mathbf{p}}},\tag{B1}
\]
and therefore,
\[
\hat{\mathrm{\mathbf{S}}}_{[2]}\cdot\frac{\hat{\mathrm{\mathbf{p}}}_{[2]}}{\left\Vert \mathrm{\mathbf{p}}_{[2]}\right\Vert }\,(\varphi_{1})_{\mathrm{\mathbf{p}}}=\mathrm{sgn}(E)\,\frac{\hbar}{2}\,\hat{1}_{2}(\varphi_{1})_{\mathrm{\mathbf{p}}}\,,\quad\hat{\mathrm{\mathbf{S}}}_{[2]}\cdot\frac{\hat{\mathrm{\mathbf{p}}}_{[2]}}{\left\Vert \mathrm{\mathbf{p}}_{[2]}\right\Vert }\,(\varphi_{2})_{\mathrm{\mathbf{p}}}=-\mathrm{sgn}(E)\,\frac{\hbar}{2}\,\hat{1}_{2}(\varphi_{2})_{\mathrm{\mathbf{p}}},\tag{B2}
\]
where $\hat{\lambda}_{[2]}\equiv\hat{\boldsymbol{\sigma}}\cdot\hat{\mathrm{\mathbf{p}}}_{[2]}/\left\Vert \mathrm{\mathbf{p}}_{[2]}\right\Vert =\hat{\mathrm{\mathbf{S}}}_{[2]}\cdot\hat{\mathrm{\mathbf{p}}}_{[2]}/\frac{\hbar}{2}\left\Vert \mathrm{\mathbf{p}}_{[2]}\right\Vert $
and $\left\Vert \mathrm{\mathbf{p}}_{[2]}\right\Vert =\left\Vert \mathrm{\mathbf{p}}\right\Vert $.
Thus, the eigenvalues of the operators $\hat{\lambda}$ and $\hat{\lambda}_{[2]}$
only indicate whether the direction of the spin of the particle in
question is parallel or antiparallel to its respective momentum; however,
all of these eigenvalues are also dependent on the sign of the energy.

Let us now introduce the so-called (Hermitian) classical velocity
operator $\hat{\mathrm{v}}_{\mathrm{cl}}\equiv c^{2}\hat{\mathrm{\mathbf{p}}}\hat{\mathrm{E}}^{-1}$
(which corresponds to the formula of classical relativistic mechanics
that provides the velocity as a function of momentum and energy),
where $\hat{\mathrm{E}}$ is the Dirac Hamiltonian operator \cite{RefZZ}.
Clearly, if $\hat{\mathrm{\mathrm{\mathbf{v}}}}_{\mathrm{cl}}$ acts
on the Dirac plane-wave solution $\Psi_{\mathrm{\mathbf{p}}}$, one
obtains the eigenvalue $\mathrm{\mathrm{\mathbf{v}}}_{\mathrm{cl}}=c^{2}\mathrm{\mathbf{p}}/E$,
i.e., $\mathrm{\mathrm{\mathbf{v}}}_{\mathrm{cl}}=\mathrm{sgn}(E)\, c\,\mathrm{\mathbf{p}}/\left\Vert \mathrm{\mathbf{p}}\right\Vert $
($\Rightarrow\left\Vert \mathrm{\mathrm{\mathbf{v}}}_{\mathrm{cl}}\right\Vert =c$,
as expected). Then, we can use these results to write the relations
in (B1) and (B2) such that they do not depend on the sign of the energy,
that is, 
\[
\hat{\mathrm{\mathbf{S}}}\cdot\frac{\hat{\mathrm{\mathrm{\mathbf{v}}}}_{\mathrm{cl}}}{c}\,(\Psi_{+})_{\mathrm{\mathbf{p}}}=\frac{\hbar}{2}\,\hat{1}_{4}(\Psi_{+})_{\mathrm{\mathbf{p}}}\,,\quad\hat{\mathrm{\mathbf{S}}}\cdot\frac{\hat{\mathrm{\mathrm{\mathbf{v}}}}_{\mathrm{cl}}}{c}\,(\Psi_{-})_{\mathrm{\mathbf{p}}}=-\frac{\hbar}{2}\,\hat{1}_{4}(\Psi_{-})_{\mathrm{\mathbf{p}}},\tag{B3}
\]
and 
\[
\hat{\mathrm{\mathbf{S}}}_{[2]}\cdot\frac{(\hat{\mathrm{\mathrm{\mathbf{v}}}}_{\mathrm{cl}})_{[2]}}{c}\,(\varphi_{1})_{\mathrm{\mathbf{p}}}=\frac{\hbar}{2}\,\hat{1}_{2}(\varphi_{1})_{\mathrm{\mathbf{p}}}\,,\quad\hat{\mathrm{\mathbf{S}}}_{[2]}\cdot\frac{(\hat{\mathrm{\mathrm{\mathbf{v}}}}_{\mathrm{cl}})_{[2]}}{c}\,(\varphi_{2})_{\mathrm{\mathbf{p}}}=-\frac{\hbar}{2}\,\hat{1}_{2}(\varphi_{2})_{\mathrm{\mathbf{p}}},\tag{B4}
\]
respectively (where $(\hat{\mathrm{\mathrm{\mathbf{v}}}}_{\mathrm{cl}})_{[2]}=\mathrm{sgn}(E)\, c\,\hat{\mathrm{\mathbf{p}}}_{[2]}/\left\Vert \mathrm{\mathbf{p}}_{[2]}\right\Vert $
and $\hat{\mathrm{\mathrm{\mathbf{v}}}}_{\mathrm{cl}}=\mathrm{diag}((\hat{\mathrm{\mathrm{\mathbf{v}}}}_{\mathrm{cl}})_{[2]},(\hat{\mathrm{\mathrm{\mathbf{v}}}}_{\mathrm{cl}})_{[2]})$).
In this way, the eigenvalues of the operators $\hat{\mathrm{\mathbf{S}}}\cdot\hat{\mathrm{\mathrm{\mathbf{v}}}}_{\mathrm{cl}}/c$
and $\hat{\mathrm{\mathbf{S}}}_{[2]}\cdot(\hat{\mathrm{\mathrm{\mathbf{v}}}}_{\mathrm{cl}})_{[2]}/c$
indicate whether the direction of the spin of the particle in question
is parallel or antiparallel to the movement of the particle. For example,
the spin of the 3D Weyl particle described by $(\varphi_{1})_{\mathrm{\mathbf{p}}}$
is always parallel to its direction of motion, but the spin of the
3D Weyl particle described by $(\varphi_{2})_{\mathrm{\mathbf{p}}}$
is always antiparallel to its direction of motion.

As we have seen, the eigenstates of the operator $\hat{\mathrm{p}}/\left|\mathrm{p}\right|$
in (1+1) dimensions, $(\Psi_{+})_{\mathrm{p}}=\left[\,(\varphi_{1})_{\mathrm{p}}\;\,0\,\right]^{\mathrm{T}}$
and $(\Psi_{-})_{\mathrm{p}}=\left[\,0\;\,(\varphi_{2})_{\mathrm{p}}\,\right]^{\mathrm{T}}$,
comply with relations that depend on the sign of the energy, namely,
\[
\frac{\hat{\mathrm{p}}}{\left|\mathrm{p}\right|}(\Psi_{+})_{\mathrm{p}}=\mathrm{sgn}(E)\hat{1}_{2}(\Psi_{+})_{\mathrm{p}}\,,\quad\frac{\hat{\mathrm{p}}}{\left|\mathrm{p}\right|}(\Psi_{-})_{\mathrm{p}}=-\mathrm{sgn}(E)\hat{1}_{2}(\Psi_{-})_{\mathrm{p}},\tag{B5}
\]
from which similar relations for $(\varphi_{1})_{\mathrm{p}}$ and
$(\varphi_{2})_{\mathrm{p}}$ are immediately obtained, namely,
\[
\frac{\hat{\mathrm{p}}_{[1]}}{\left|\mathrm{p}_{[1]}\right|}(\varphi_{1})_{\mathrm{p}}=\mathrm{sgn}(E)(\varphi_{1})_{\mathrm{p}}\,,\quad\frac{\hat{\mathrm{p}}_{[1]}}{\left|\mathrm{p}_{[1]}\right|}(\varphi_{2})_{\mathrm{p}}=-\mathrm{sgn}(E)(\varphi_{2})_{\mathrm{p}},\tag{B6}
\]
where $\hat{\mathrm{p}}/\left|\mathrm{p}\right|=\hat{\mathrm{p}}_{[1]}\hat{1}_{2}/\left|\mathrm{p}_{[1]}\right|$
and $\left|\mathrm{p}_{[1]}\right|=\left|\mathrm{p}\right|$. Clearly,
the operators $\hat{\mathrm{p}}/\left|\mathrm{p}\right|$ and $\hat{\mathrm{\mathbf{S}}}\cdot\hat{\mathrm{\mathbf{p}}}/\left\Vert \mathrm{\mathbf{p}}\right\Vert $,
as well as $\hat{\mathrm{p}}_{[1]}/\left|\mathrm{p}_{[1]}\right|$
and $\hat{\mathrm{\mathbf{S}}}_{[2]}\cdot\hat{\mathrm{\mathbf{p}}}_{[2]}/\left\Vert \mathrm{\mathbf{p}}_{[2]}\right\Vert $,
have a certain similarity (when acting on their respective chiral
plane-wave eigenstates). The Dirac plane-wave $\Psi_{\mathrm{p}}$
is also an eigensolution of the (Hermitian) classical velocity operator
$\hat{\mathrm{v}}_{\mathrm{cl}}\equiv c^{2}\hat{\mathrm{p}}\hat{\mathrm{E}}^{-1}$
and has eigenvalue $\mathrm{v}_{\mathrm{cl}}=c^{2}\mathrm{p}/E=\mathrm{sgn}(E)\, c\,\mathrm{p}/\left|\mathrm{p}\right|$
($\hat{\mathrm{E}}\,(=\hat{\mathrm{h}})$ is the Dirac Hamiltonian
operator in (1+1) dimensions). This fact allows us to write the relations
in (B5) and (B6) in a form independent of the energy sign, namely,
\[
\frac{\hat{\mathrm{v}}_{\mathrm{cl}}}{c}(\Psi_{+})_{\mathrm{p}}=\hat{1}_{2}(\Psi_{+})_{\mathrm{p}}\,,\quad\frac{\hat{\mathrm{v}}_{\mathrm{cl}}}{c}(\Psi_{-})_{\mathrm{p}}=-\hat{1}_{2}(\Psi_{-})_{\mathrm{p}},\tag{B7}
\]
and
\[
\frac{(\hat{\mathrm{v}}_{\mathrm{cl}})_{[1]}}{c}(\varphi_{1})_{\mathrm{p}}=(\varphi_{1})_{\mathrm{p}}\,,\quad\frac{(\hat{\mathrm{v}}_{\mathrm{cl}})_{[1]}}{c}(\varphi_{2})_{\mathrm{p}}=-(\varphi_{2})_{\mathrm{p}},\tag{B8}
\]
respectively (where $(\hat{\mathrm{v}}_{\mathrm{cl}})_{[1]}=\mathrm{sgn}(E)\, c\,\hat{\mathrm{p}}_{[1]}/\left|\mathrm{p}_{[1]}\right|$
and $\hat{\mathrm{v}}_{\mathrm{cl}}=(\hat{\mathrm{v}}_{\mathrm{cl}})_{[1]}\hat{1}_{2}$).
Clearly, the eigenvalues of the operators $\hat{\mathrm{v}}_{\mathrm{cl}}/c$
and $(\hat{\mathrm{v}}_{\mathrm{cl}})_{[1]}/c$ indicate whether the
particle in question, whether it is a 1D Dirac particle or a 1D Weyl
particle, actually moves to the right or to the left. For example,
the 1D Weyl particle described by $(\varphi_{1})_{\mathrm{p}}$ always
moves to the right (left), but the 1D Weyl particle described by $(\varphi_{2})_{\mathrm{p}}$
moves to the left (right). 

\subsection{On the boundary conditions for the Weyl equations}

\noindent We have obtained the most general families of boundary conditions
for the (time-dependent) Weyl equations given in Eq. (14) (i.e., in
(3+1) dimensions), where the (self-adjoint) Weyl Hamiltonian operators
present are precisely the operators $\hat{\mathrm{H}}_{a,j}$ given
in Eq. (17). Each of the three families of boundary conditions (labeled
by $j=1,2,3$ and given in Eqs. (31), (33) and (35)) is parametrized
by a unitary $2\times2$ matrix, that is, by $2^{2}=4$ real parameters.
A feasible parametrization for these unitary matrices, for example,
for the matrix $\hat{A}_{1}$ in Eq. (31), is given by 
\[
\hat{A}_{1}=\exp(\mathrm{i}\mu)\left[\begin{array}{cc}
m_{0}-\mathrm{i}m_{3} & -m_{2}-\mathrm{i}m_{1}\\
m_{2}-\mathrm{i}m_{1} & m_{0}+\mathrm{i}m_{3}
\end{array}\right],\tag{C1}
\]
where $\mu\in[0,\pi)$, and real quantities $m_{0}$, $m_{1}$, $m_{2}$
and $m_{3}$, satisfy $(m_{0})^{2}+(m_{1})^{2}+(m_{2})^{2}+(m_{3})^{2}=1$
(but also $\det(\hat{A}_{1})=\exp(\mathrm{i}2\mu)$) \cite{RefZZZ}.
For other interesting examples of Hamiltonians operators whose self-adjoint
extensions (or sets of general boundary conditions) are characterized
in terms of unitary matrices, see Refs. \cite{RefZZZZ,RefZZZZZ}.

On the other hand, all boundary conditions that are part of each of
these three families of self-adjoint boundary conditions cancel the
boundary term in Eq. (18), which implies that 
\[
\left.c\left[\,\varphi_{a,j}^{\dagger}\hat{\sigma}_{j}\varphi_{a,j}\,\right]\right|_{0}^{\ell}\equiv\left.\left[\, J_{a,j}\,\right]\right|_{0}^{\ell}=0\quad\Rightarrow\quad J_{a,j}(x^{j}=\ell,t)=J_{a,j}(x^{j}=0,t),\tag{C2}
\]
where $J_{a,j}=J_{a,j}(x^{j},t)$ is the probability current density
\cite{RefK}. Thus, all of the self-adjoint boundary conditions lead
to the equality of $J_{a,j}$ at the ends of the box. Within each
general family of boundary conditions, there are boundary conditions
that simply cancel the probability current density at these extremes;
they are called confining boundary conditions. For example, the following
confining boundary conditions for the Weyl Hamiltonian $\hat{\mathrm{H}}_{a,1}$
are contained in Eq. (31): $\varphi_{a}^{\mathrm{t}}(x=\ell,t)=\varphi_{a}^{\mathrm{t}}(x=0,t)=0$
($\hat{A}_{1}=-\hat{1}_{2}$), i.e., the upper component of the wave
function $\varphi_{a,1}\equiv\varphi_{a}$ can satisfy the Dirichlet
boundary condition; $\varphi_{a}^{\mathrm{b}}(x=\ell,t)=\varphi_{a}^{\mathrm{b}}(x=0,t)=0$
($\hat{A}_{1}=+\hat{1}_{2}$), i.e., the lower component of the wave
function $\varphi_{a,1}\equiv\varphi_{a}$ can also satisfy the Dirichlet
boundary condition. However, the entire two-component Weyl wave function
$\varphi_{a,1}\equiv\varphi_{a}$ does not support this boundary condition
at the walls of the box, i.e., the latter is not contained in Eq.
(31). This result is also fulfilled by the Dirac wave function \cite{RefZZZZZZ}.
Likewise, there are also boundary conditions that do not cancel $J_{a,j}$
at the ends of the box; they are called non-confining boundary conditions.
For example, the following non-confining boundary conditions for the
Weyl Hamiltonian $\hat{\mathrm{H}}_{a,1}$ are also contained in Eq.
(31): $\varphi_{a}^{\mathrm{t}}(x=\ell,t)=\varphi_{a}^{\mathrm{t}}(x=0,t)$
and $\varphi_{a}^{\mathrm{b}}(x=\ell,t)=\varphi_{a}^{\mathrm{b}}(x=0,t)$
($\hat{A}_{1}=+\hat{\sigma}_{x}$), i.e., the wave function $\varphi_{a,1}\equiv\varphi_{a}$
can satisfy the periodic boundary condition; $\varphi_{a}^{\mathrm{t}}(x=\ell,t)=-\varphi_{a}^{\mathrm{t}}(x=0,t)$
and $\varphi_{a}^{\mathrm{b}}(x=\ell,t)=-\varphi_{a}^{\mathrm{b}}(x=0,t)$
($\hat{A}_{1}=-\hat{\sigma}_{x}$), i.e., the wave function $\varphi_{a,1}\equiv\varphi_{a}$
can also satisfy the antiperiodic boundary condition. 

As was noted in section III, the (time-dependent) Weyl equations with
the (self-adjoint) Hamiltonian operators $\hat{\mathrm{H}}_{a,1}$
and $\hat{\mathrm{H}}_{a,3}$ can provide purely real-valued solutions.
Thus, if we impose on the respective wave functions the reality condition,
these wave functions and their respective complex conjugates must
satisfy the same boundary conditions, in which case the unitary matrices
$\hat{A}_{1}$ and $\hat{A}_{3}$ must each be orthogonal. For example,
in this case, the unitary matrix $\hat{A}_{1}$ in Eq. (C1) takes
the form
\[
\hat{A}_{1}=\left[\begin{array}{cc}
m_{0} & -m_{2}\\
m_{2} & m_{0}
\end{array}\right],\tag{C3}
\]
where $(m_{0})^{2}+(m_{2})^{2}=1$, and therefore, $\det(\hat{A}_{1})=+1$
(because $m_{1}=m_{3}=0$ and $\mu=0$). Likewise, $\hat{A}_{1}$
in Eq. (C1) can also take the form 
\[
\hat{A}_{1}=\left[\begin{array}{cc}
m_{3} & m_{1}\\
m_{1} & -m_{3}
\end{array}\right],\tag{C4}
\]
where $(m_{1})^{2}+(m_{3})^{2}=1$, and therefore, $\det(\hat{A}_{1})=-1$
(because $m_{0}=m_{2}=0$ and $\mu=\pi/2$) \cite{RefZZZ}. Contrarily,
the (time-dependent) Weyl equation with the (self-adjoint) Hamiltonian
operator $\hat{\mathrm{H}}_{a,2}$ cannot provide real-valued solutions;
thus, the corresponding wave functions support any boundary condition
included in Eq. (33). 

It is a known fact that families of boundary conditions for relativistic
(and non-relativistic) Hamiltonian operators that describe a quantum
particle inside a finite interval, and on the entire real line except
the finite interval, are similar (in the latter case, if the interval
is very small, we have the real line with a single point excluded
-- a hole --, and a particle in this kind of system can be modeled
through proper boundary conditions only, and with potentials only
-- with singular potentials and with smooth potentials --). For example,
in Ref. \cite{RefZZZZZZZ}, 1D Dirac point interactions were recently
modeled with a two-parameter potential that has a scalar part and
an electrostatic part, and each is essentially a Dirac delta function.
Likewise, in Ref. \cite{RefZZZZZZZZ}, some self-adjoint extensions
of the 1D Dirac Hamiltonian operator for massive ``1D-fermions''
confined to the interval $\Omega=[-L/2,+L/2]$ were recently studied
(the latter results can be conveniently extended to the region $\mathbb{R}-\Omega$,
which the authors call the dual geometry of $\Omega$). Essentially,
the results in Ref. \cite{RefZZZZZZZZ} were applied to the calculation
of the so-called Casimir energy of a massive Dirac field confined
in a 1D finite filament (of length $L$). Precisely, the most general
family of (self-adjoint) boundary conditions for the Dirac Hamiltonian
operator (in Eq. (3) of Ref. \cite{RefZZZZZZZZ}) is given by 
\[
\left[\begin{array}{c}
\psi_{1}(x=-L/2)\\
\psi_{2}(x=+L/2)
\end{array}\right]=\hat{\mathbb{U}}\hat{\gamma}_{0}\left[\begin{array}{c}
\psi_{1}(x=+L/2)\\
\psi_{2}(x=-L/2)
\end{array}\right]\tag{C5}
\]
(see Eq. (23) in Ref. \cite{RefZZZZZZZZ}), where $\hat{\mathbb{U}}\hat{\gamma}_{0}$
is a unitary matrix (this is because $\hat{\mathbb{U}}$ and $\hat{\gamma}_{0}$
are also unitary matrices). The latter family of boundary conditions
is similar to the most general family of boundary conditions for the
self-adjoint Weyl operator $\hat{\mathrm{H}}_{a,3}$. In fact, the
set of boundary conditions for $\hat{\mathrm{H}}_{a,3}$ is given
in Eq. (35), but it can also be written as the set in Eq. (C5), namely,
\[
\left[\begin{array}{c}
\varphi_{a}^{\mathrm{t}}(z=0)\\
\varphi_{a}^{\mathrm{b}}(z=\ell)
\end{array}\right]=(\hat{A}_{3}\hat{\sigma}_{x})^{-1}\left[\begin{array}{c}
\varphi_{a}^{\mathrm{t}}(z=\ell)\\
\varphi_{a}^{\mathrm{b}}(z=0)
\end{array}\right],\tag{C6}
\]
where $(\hat{A}_{3}\hat{\sigma}_{x})^{-1}$ is a unitary matrix (this
is because $\hat{A}_{3}$ and $\hat{\sigma}_{x}$ are also unitary
matrices). For simplicity, the variable $t$ was dropped from the
functions $\varphi_{a}^{\mathrm{t}}$ and $\varphi_{a}^{\mathrm{b}}$
in Eq. (C6). Again, all the boundary conditions for the (massive)
1D Dirac particle in the box are also valid for the (massless) 3D
Weyl particle in the box. 

Finally, as might be expected, within the family given in Eq. (C5),
and in Eq. (C6), we also have the boundary condition commonly used
in the so-called MIT bag model for hadronic structure (certainly,
in its one-dimensional version) \cite{RefZZZZZZZZZ}. This confining
boundary condition can be obtained from Eq. (C5), where the matrix
$\hat{\mathbb{U}}\hat{\gamma}_{0}$ satisfies the relation $\hat{\mathbb{U}}\hat{\gamma}_{0}\,\hat{\alpha}+\hat{\alpha}\,\hat{\mathbb{U}}\hat{\gamma}_{0}=\hat{0}_{2}$
(see Eq. (27) in Ref. \cite{RefZZZZZZZZ}), by setting $\theta=-\pi/2$
and $\eta=0$, and therefore, $\hat{\mathbb{U}}\hat{\gamma}_{0}=(-1)^{r}(-\mathrm{i})\hat{\sigma}_{x}$.
Thus, one obtains 
\[
\psi_{1}(-L/2)=(-1)^{r}(-\mathrm{i})\,\psi_{2}(-L/2),\quad\psi_{1}(+L/2)=(-1)^{r}(+\mathrm{i})\,\psi_{2}(+L/2).\tag{C7}
\]
Interestingly, the latter boundary condition can also be written explicitly
in a form independent of the particular choice of the gamma matrices
(and, as is known, also in a Lorentz-invariant way), namely, 
\[
\mathrm{i}\, n_{\mu}\hat{\gamma}^{\mu}\psi=\psi\;\mathrm{at}\; x=-L/2\;\mathrm{and}\; x=+L/2,\tag{C8}
\]
where $n^{\mu}=(0,-1)$ at $x=-L/2$, and $n^{\mu}=(0,+1)$ at $x=+L/2$
(i.e., the unit two-vector normal to the surface of the box is pointing
outward from the wall). In addition, we have that $\hat{\gamma}^{0}=(-1)^{r}\hat{\sigma}_{x}$
and $\hat{\gamma}^{1}=\hat{\gamma}^{0}\hat{\alpha}=\hat{\gamma}^{0}\hat{\sigma}_{z}=(-1)^{r}(-\mathrm{i})\hat{\sigma}_{y}$
(see Eq. (26) in Ref. \cite{RefZZZZZZZZ}). Similarly, the usual MIT
bag boundary condition can be obtained from the family in Eq. (C6)
by setting $(\hat{A}_{3}\hat{\sigma}_{x})^{-1}=(-\mathrm{i})\hat{\sigma}_{x}$.
Thus, one obtains $\varphi_{a}^{\mathrm{t}}(z=0)=(-\mathrm{i})\varphi_{a}^{\mathrm{b}}(z=0)$,
$\varphi_{a}^{\mathrm{t}}(z=\ell)=(+\mathrm{i})\varphi_{a}^{\mathrm{b}}(z=\ell)$. 

On the other hand, in (1+1) dimensions, the most general family of
self-adjoint boundary conditions for each of the (time-dependent)
Weyl equations given in Eq. (36) is characterized by a phase, i.e.,
by a single real parameter. All boundary conditions present in these
two families of boundary conditions cancel the boundary term in Eq.
(39), which implies that
\[
\left.\left[\,\varphi_{a}^{*}\varphi_{a}\,\right]\right|_{0}^{\ell}\equiv\left.\left[\,\varrho_{a}\,\right]\right|_{0}^{\ell}=0\quad\Rightarrow\quad\varrho_{a}(x=\ell,t)=\varrho_{a}(x=0,t),\tag{C9}
\]
where $\varrho_{a}=\varrho_{a}(x,t)$ is the probability density.
In this case, each Weyl equation leads to an atypical continuity equation,
in which the probability density is precisely proportional to the
probability current density, namely, $\partial(\varphi_{a}^{*}\varphi_{a})/\partial t+(-1)^{a-1}\partial(c\,\varphi_{a}^{*}\varphi_{a})/\partial x=0$.
With that said, it is clear that all boundary conditions within the
two one-parametric families of boundary conditions are non-confining
boundary conditions, i.e., none of them can cancel the probability
current density at the ends of the box.

\begin{acknowledgments}
\noindent I thank Valedith Cusati, my wife, for all her support.
\end{acknowledgments}

\end{document}